\documentclass{iopjournal}
\usepackage{graphicx} 
\usepackage[backend=biber,citestyle=phys,style=numeric,sorting=none]{biblatex}
\usepackage{qm}
\usepackage{tikz}
\usepackage{subcaption}
\usepackage{qcircuit}
\usepackage{ragged2e}

\newcommand{\twoline}[2]{{\text{\begin{tabular}{c}#1\\#2\end{tabular}}}}

\newcommand{\reuploadlayer}{\Qcircuit @C=1em @R=1.5em {
		\lstick{q_0} 
		& \gate{R_X(x_0^{(j)})} & \gate{R_Y(y_0^{(j)})} & \gate{R_Z(z_0^{(j)})} & \gate{R(w_{0}^{(j)})} & \ctrl{1} & \qw \\
		\lstick{q_1} 
		& \gate{R_X(x_1^{(j)})} & \gate{R_Y(y_1^{(j)})} & \gate{R_Z(z_1^{(j)})} & \gate{R(w_{1}^{(j)})} & \ctrl{-1} & \ctrl{1} \\
		\lstick{q_2} 
		& \gate{R_X(x_2^{(j)})} & \gate{R_Y(y_2^{(j)})} & \gate{R_Z(z_2^{(j)})} & \gate{R(w_{2}^{(j)})} & \qw & \ctrl{-1} \\
}}

\newcommand{\finalmeasurementlayer}{
	\Qcircuit @C=1em @R=1.5em {
		\lstick{q_0} 
		& \gate{R_X(x_0^n + w_0^n)} & \gate{R_Y(y_0^n + w_0^n)} & \gate{R_Z(z_0^n + w_0^n)} & \meter \\
		\lstick{q_1} 
		& \gate{R_X(x_1^n + w_1^n)} & \gate{R_Y(y_1^n + w_1^n)} & \gate{R_Z(z_1^n + w_1^n)} & \meter \\
		\lstick{q_2} 
		& \gate{R_X(x_2^n + w_2^n)} & \gate{R_Y(y_2^n + w_2^n)} & \gate{R_Z(z_2^n + w_2^n)} & \meter \\
}}

\newcommand{\encoder}[2][Image]{
	\begin{tikzpicture}[grow=right]
		\node[circle,draw,] (IMGin) at (-2, 0)  {#1};

		\node[circle,draw,] (A0) at (-1, -1)  {};
		\node[circle,draw,] (A1) at (-1, 0)  {};
		\node[circle,draw] (A2) at (-1, 1)  {};
		
		\node[circle,draw,] (B0) at (0, -1)  {};
		\node[circle,draw,] (B1) at (0, 0)  {};  
		\node[circle,draw,] (B2) at (0, 1)  {};
		
		\node[circle,draw,] (C0) at (1, -1)  {};
		\node[circle,draw,] (C1) at (1, 0)  {};  
		\node[circle,draw,] (C2) at (1, 1)  {};
		
		\node[rectangle,draw,] (Qs) at (2, 0)  {#2};

		\path [->] (IMGin) edge node[near start,above] {} (A0);
		\path [->] (IMGin) edge node[near start,below] {} (A1);
		\path [->] (IMGin) edge node[near start,above] {} (A2);
		
		\path [-] (A0) edge node[near start,above] {} (B0);
		\path [-] (A1) edge node[near start,below] {} (B0);
		\path [-] (A2) edge node[near start,above] {} (B0);
		\path [-] (A0) edge node[near start,above] {} (B1);
		\path [-] (A1) edge node[near start,below] {} (B1);
		\path [-] (A2) edge node[near start,above] {} (B1);
		\path [-] (A0) edge node[near start,above] {} (B2);
		\path [-] (A1) edge node[near start,below] {} (B2);
		\path [-] (A2) edge node[near start,above] {} (B2);
		
		\path [-] (B0) edge node[near start,above] {} (C0);
		\path [-] (B1) edge node[near start,below] {} (C0);
		\path [-] (B2) edge node[near start,above] {} (C0);
		
		\path [-] (B0) edge node[near start,above] {} (C1);
		\path [-] (B1) edge node[near start,below] {} (C1);
		\path [-] (B2) edge node[near start,above] {} (C1);
		
		\path [-] (B0) edge node[near start,above] {} (C2);
		\path [-] (B1) edge node[near start,below] {} (C2);
		\path [-] (B2) edge node[near start,above] {} (C2);

		\path [->] (C0) edge node[near start,above] {} (Qs);
		\path [->] (C1) edge node[near start,below] {} (Qs);
		\path [->] (C2) edge node[near start,above] {} (Qs);            
	\end{tikzpicture}
}

\newcommand{\decoder}{
	\begin{tikzpicture}[grow=right]
		\node[rectangle,draw,] (Qm) at (9, 0)        {\twoline{Measurement}{Data}};
		
		\node[circle,draw,] (DA0) at (11, -1)  {};
		\node[circle,draw,] (DA1) at (11, 0)  {};
		\node[circle,draw] (DA2) at (11, 1)  {};
		
		\node[circle,draw,] (DB0) at (12, -1)  {};
		\node[circle,draw,] (DB1) at (12, 0)  {};  
		\node[circle,draw,] (DB2) at (12, 1)  {};
		
		\node[circle,draw,] (DC0) at (13, -1)  {};
		\node[circle,draw,] (DC1) at (13, 0)  {};  
		\node[circle,draw,] (DC2) at (13, 1)  {};
		
		\node[circle,draw,] (IMGout) at (15, 0)  {\begin{tabular}{c}
				Image   \\
				Reconstructed
		\end{tabular} };
		
		\path [->] (Qm) edge node[near start,above] {} (DA0);
        \path [->] (Qm) edge node[near start,above] {} (DA1);
        \path [->] (Qm) edge node[near start,above] {} (DA2);
		
		\path [-] (DA0) edge node[near start,above] {} (DB0);
		\path [-] (DA1) edge node[near start,below] {} (DB0);
		\path [-] (DA2) edge node[near start,above] {} (DB0);
		\path [-] (DA0) edge node[near start,above] {} (DB1);
		\path [-] (DA1) edge node[near start,below] {} (DB1);
		\path [-] (DA2) edge node[near start,above] {} (DB1);
		\path [-] (DA0) edge node[near start,above] {} (DB2);
		\path [-] (DA1) edge node[near start,below] {} (DB2);
		\path [-] (DA2) edge node[near start,above] {} (DB2);
		
		\path [-] (DB0) edge node[near start,above] {} (DC0);
		\path [-] (DB1) edge node[near start,below] {} (DC0);
		\path [-] (DB2) edge node[near start,above] {} (DC0);
		
		\path [-] (DB0) edge node[near start,above] {} (DC1);
		\path [-] (DB1) edge node[near start,below] {} (DC1);
		\path [-] (DB2) edge node[near start,above] {} (DC1);
		
		\path [-] (DB0) edge node[near start,above] {} (DC2);
		\path [-] (DB1) edge node[near start,below] {} (DC2);
		\path [-] (DB2) edge node[near start,above] {} (DC2);
		
		\path [->] (DC0) edge node[near start,above] {} (IMGout);
		\path [->] (DC1) edge node[near start,below] {} (IMGout);
		\path [->] (DC2) edge node[near start,above] {} (IMGout);
	\end{tikzpicture}	
}

\newcommand{\postprocessinglayer}[1][Probs]{
	\begin{tikzpicture}[grow=right]
		\node[rectangle,draw] (Probs) at (0, 0) {#1};
		\node[circle,draw] (Node1) at (2, 0.5) {};
		\node[circle,draw] (Node2) at (2, -0.5) {};
		\node[circle,draw] (Softmax) at (4, 0) {Softmax};
		
		\draw[->] (Probs) -- (Node1);
		\draw[->] (Probs) -- (Node2);
		\draw[->] (Node1) -- (Softmax);
		\draw[->] (Node2) -- (Softmax);
	\end{tikzpicture}
	
}

\newcommand{\amplitudeembedding}[1][$x$]{
	\Qcircuit @C=0.9em @R=1.5em {
		\lstick{q_0} & \multigate{2}{\text{AmplitudeEmbedding(#1)}} & \qw \\
		\lstick{q_1} & \ghost{\text{AmplitudeEmbedding(#1)}} & \qw \\
		\lstick{q_2} & \ghost{\text{AmplitudeEmbedding(#1)}} & \qw \\
	}
	
}

\newcommand{\amplitudeclassifier}{
	\(
	\Qcircuit @C=0.9em @R=1.5em {
		\lstick{q_0} 
		& \multigate{2}{\text{AmplitudeEmbedding($x$)}} 
		& \multigate{2}{\text{Strongly Entangled Layer~$(\theta)_1$}} 
		&\qw & \cdots &
		& \multigate{2}{\text{Strongly Entangled Layer~$(\theta)_i$}} 
		&\qw & \cdots &
		& \multigate{2}{\text{Strongly Entangled Layer~$(\theta)_n$}} 
		& \meter \\
		\lstick{q_1} 
		& \ghost{\text{AmplitudeEmbedding($x$)}} 
		& \ghost{\text{Strongly Entangled Layer~$(\theta)_1$}} 
		&\qw & \cdots &
		& \ghost{\text{Strongly Entangled Layer~$(\theta)_i$}} 
		&\qw & \cdots &
		& \ghost{\text{Strongly Entangled Layer~$(\theta)_n$}} 
		& \meter \\
		\lstick{q_2} 
		& \ghost{\text{AmplitudeEmbedding($x$)}} 
		& \ghost{\text{Strongly Entangled Layer~$(\theta)_1$}} 
		&\qw & \cdots &
		& \ghost{\text{Strongly Entangled Layer~$(\theta)_i$}} 
		&\qw & \cdots &
		& \ghost{\text{Strongly Entangled Layer~$(\theta)_n$}} 
		& \meter \\
	}
	\)
	
}

\newcommand{\stronglyentangledlayercircuit}{
	
	\Qcircuit @C=0.9em @R=1.5em {
		\lstick{q_0} 
		& \gate{RX(\alpha_{0})} & \gate{RY(\beta_{0})} & \gate{RZ(\gamma_{0})} & \ctrl{1} & \qw & \ctrl{2} & \qw &\qw  & \qw \\
		\lstick{q_1} 
		& \gate{RX(\alpha_{1})} & \gate{RY(\beta_{1})} & \gate{RZ(\gamma_{1})} & \ctrl{-1} & \ctrl{1} & \qw & \qw & \qw & \qw  \\
		\lstick{q_2} 
		& \gate{RX(\alpha_{2})} & \gate{RY(\beta_{2})} & \gate{RZ(\gamma_{2})} & \qw & \ctrl{-1} & \ctrl{-2} & \qw & \qw & \qw  \\
	}
	
}

\begin{document}
\justifying

\articletype{paper} 

\title{Tailor Made Embeddings for Quantum Machine Learning}

\author{Aldo Lamarre$^{1,2,*}$\orcid{0000-0000-0000-0000} and Dominik \v{S}afr\'{a}nek$^{2,3,\dag}$\orcid{0000-0002-6861-395X}}

\affil{$^1$Department of Electrical Engineering, KAIST, Daejeon, South Korea}

\affil{$^2$Center for Theoretical Physics of Complex Systems, Institute for Basic Science (IBS), Daejeon - 34126, Korea}

\affil{$^3$Faculty of Mathematics and Physics, Charles University, Ke Karlovu 3, 121 16 Praha 2, Czech Republic}

\affil{$^*$\hyperlink{lamarral@kaist.ac.kr}{lamarral@kaist.ac.kr}}

\affil{$^\dag$\hyperlink{dominik.safranek@matfyz.cuni.cz}{dominik.safranek@matfyz.cuni.cz}}

\keywords{quantum machine learning, quantum autoencoder, image processing}

\begin{abstract}
\justify
Autoencoders transformed classical machine learning by solving the curse of dimensionality, enabling principled weight initialization and learning compact, structured representations. 
In this work, we extend this paradigm to quantum machine learning by introducing a variational autoencoder framework that learns task-specific quantum embeddings of classical data. We demonstrate that high-dimensional datasets, including ImageNet, can be compressed into a 13-qubit quantum representation while remaining reconstructable through a learned decoder.
On MNIST (3 vs.\ 5), our approach achieves 98.5\% validation accuracy using a circuit-centric quantum classifier, within 1.2 percentage points of a classical neural network baseline (99.7\%) and more than 30 percentage points above a naive amplitude-embedding approach. Unlike amplitude embeddings, which require full quantum state tomography for recovery, or angle embeddings, which generally rely on circuit inversion under restrictive assumptions, the proposed framework reconstructs the original data from only a polynomial number of measurements.
The framework was further validated on IBM quantum hardware, confirming that the learned embeddings remain stable and reconstructable under real device noise. 
\end{abstract}

\section{Introduction}

Quantum machine learning models currently underperform well-engineered classical machine learning models in practical applications~\cite{bowles2024betterclassicalsubtleart}. This performance gap can be attributed not only to hardware limitations, but also to the relative lack of mature machine learning engineering practices and training heuristics, sometimes informally referred to as the “black arts” of classical machine learning.

Quantum machine learning still has considerable room for development in terms of training heuristics and optimization strategies, particularly when compared with the extensive engineering practices that have been established in classical machine learning.

In theory, several quantum variational algorithms~\cite{bowles2024betterclassicalsubtleart,1804.00633,P_rez_Salinas_2020,} have universal approximation results, which guaranties that a suitable set of parameters exists in principle. This, however, does not imply that the corresponding parameters can be found efficiently through training, indicating that expressibility is not the only bottleneck. For example, practical learning and trainability problems like barren plateau \cite{McClean_2018} and data encodings \cite{lloyd2020quantumembeddingsmachinelearning} remain significant challenges.

The no free lunch theorem \cite{585893} states that no learning algorithm is universally superior across all probability distributions, indicating that any quantum advantage in machine learning algorithms can be achieved only for specific data and  tasks. Imagine a machine learning agent, i.e., a black box. This agent receives as input a quantum state produced by amplitude embeddings, defined as $\ket{\phi} =\sum_{i=0}^{n-1} \alpha_i \ket{i}$. It is tasked with reconstructing the original data. This task is non-trivial. Due to Holevo's bound, a single measurement or even a polynomial number of measurements cannot recover the information encoded in the quantum amplitudes $\alpha_i$.  To fully recover this data, the agent needs to perform full quantum state tomography, which scales exponentially in the number of qubits.  This exponential reconstruction cost places the quantum agent at a fundamental disadvantage compared to a classical agent, which can access the original data directly, and motivates our autoencoder approach. Rather than accepting exponential measurement cost as inevitable, we instead learn embeddings that are designed to be recoverable with only a polynomial number of measurements, as we describe in the following section.

\subsection{Motivation}
\label{subsec:Motivation}

The primary motivation for adopting an autoencoder framework is to enable quantum machine learning on large datasets such as ImageNet~\cite{ILSVRC15}. The limited size of both simulable and current hardware quantum computers makes it impractical to encode such datasets directly using existing quantum embeddings. Furthermore, we aim to learn embeddings that respect the constraints imposed by quantum mechanics and quantum information theory. Because current quantum models are significantly more limited than state-of-the-art classical machine learning models, providing embeddings that are incompatible with these constraints may hinder their performance. By ensuring that the embedding process is physically motivated, we seek to maximize empirical performance.

As an additional benefit, we leverage the autoencoder framework to address some of the "black arts" of machine learning, such as weight initialization and gradient flow, thereby improving the optimization of quantum models. In particular, unsupervised pre-training updates the parameters of the quantum ansatz directly within the encoder---decoder pipeline, providing a favorable initialization before downstream tasks. Furthermore, given the substantial performance gap between classical and quantum machine learning, it is important to isolate the contribution of the quantum circuit and avoid situations in which the classical autoencoder performs nearly all of the computation. By decoupling the training of the autoencoder from that of the quantum classifier, we can better assess the contribution of each component and verify that the quantum circuit provides meaningful added value.

\subsection{Problems with current embedding}
\label{subsec:CurrentLimitation}

Holevo's bound \cite{Holevo1973BoundsFT} states that for an $n$ qubit state, we can only recover at most $n$ bits of classical information at measurement per copy. In practice, current quantum machine algorithms estimate expectation values or measurement probabilities by repeated execution of the circuits to approximate the statistics.
As mentioned in the previous section \ref{subsec:Motivation}, in the case of Amplitude embedding \cite{1804.00633}, where the data is embedded in the amplitude of a quantum state, reconstructing the original data in a tractable way would contradict Holevo's bound. To fully recover the data, we need an exponential number of copies, as we need to perform full quantum state tomography. 

This bottleneck may explain the weak empirical performance of amplitude embedding in \textcite{bowles2024betterclassicalsubtleart}. Since the information is spread across exponentially many amplitudes, it may not be accessible by a quantum machine learning model.
Autoencoders \cite{UnsupervisedPretraining,vincent2008extracting,Kingma2014} can be trained to learn structured and recoverable embeddings. By selecting a tailor-made measurement scheme limited to a polynomial number of circuit evaluations, the embedding remains compatible with fundamental quantum information theory limits such as Holevo’s bound.

{\centering
\begin{table}[h]
	\centering
	\begin{tabular}{|l|c|c|l|}
		\hline
		\textbf{Encoding} & \textbf{Qubits Used} & \textbf{Runtime} & \textbf{Data Recoverability} \\
		\hline
		Basis & $O(N \cdot m)$ & $O(N \cdot m)$ & Trivial (Exact) \\
		\hline
		Angle & $O(N)$ & $O(N)$ & Polynomial (Expectation) \\
		\hline
		Data Re-uploading & $O(k), k < N$ & $O(N \cdot d)$ & Polynomial (Expectation) \\
		\hline
		Amplitude & $O(\log N)$ & $O(N)$ & Exponential (Tomography) \\
		\hline		
	\end{tabular}
	\caption{Comparison of quantum embeddings. Here, $N$ denotes the number of features, $m$ the bit precision, and $d$ the number of data re-uploading layers. Data re-uploading reduces the qubit requirement ($k$) at the cost of increasing the runtime to $O(Nd)$. Data reconcilability assumes that the final embedded state $\ket{\phi}$ is given.}
	\label{TAB:EMB}
\end{table}}


In Table~\ref{TAB:EMB}, we present the quantum embeddings considered in this work. For each embedding, we report the number of qubits required, the cost of state preparation, and the difficulty of recovering the classical information.
These embeddings are widely used in quantum machine learning \cite{bowles2024betterclassicalsubtleart,1804.00633,lloyd2020quantumembeddingsmachinelearning,Liu_2025} and illustrate different trade-offs between data compression and the tractability of data recovery. Here, tractability means that there is an efficient algorithm that can solve it in polynomial time.

\subsection{Why autoencoders}

As discussed in Section \ref{subsec:CurrentLimitation}, autoencoders solved many initial problems in classical machine learning, such as the curse of dimensionality \cite{hinton2006reducing}, weight initialization \cite{bengio2007greedy}, unsupervised pretraining \cite{UnsupervisedPretraining}, data embedding \cite{Kingma2014}, and feature learning \cite{vincent2008extracting}. 

Quantum machine learning faces similar problems, including inefficient data embeddings \cite{lloyd2020quantumembeddingsmachinelearning,}, the scalability of state preparation \cite{bowles2024betterclassicalsubtleart}, and the difficulty of training variational circuits \cite{Schuld_2019,McClean_2018}. Classical deep learning models suffered from similar underperformance before the development of modern training techniques such as improved weight initialization \cite{glorot2010understanding, he2015delving}, unsupervised pretraining \cite{UnsupervisedPretraining}, and robust feature learning \cite{vincent2008extracting}. Quantum machine learning currently lacks equivalent engineering practices — optimizer selection, initialization strategies, and embedding design remain largely ad hoc \cite{bowles2024betterclassicalsubtleart}. We propose using autoencoders to address two of these gaps simultaneously: providing a principled embedding that respects quantum information constraints, and supplying better initial parameters for variational circuits through unsupervised pretraining \cite{UnsupervisedPretraining,bengio2007greedy}. This mirrors the historical role autoencoders played in classical deep learning before modern initialization techniques made them less necessary \cite{UnsupervisedPretraining, vincent2008extracting}.

Finally, the autoencoder and the quantum model are trained independently. The autoencoder is optimized first and its parameters are fixed afterwards. Then its latent representation is fed to the quantum model. This separation allows to isolate the contribution of the classical representation learning and the quantum classifier. Doing so permits us to fairly evaluate the performance of the quantum models directly and compare them to classical models on the same encoding.

\subsection{Our Contribution}

In this paper, we present a framework to create tailor-made QML embeddings with deep variational auto-encoders \cite{Kingma2014} (Section~\ref{subsec:encoder}). The framework unifies representation learning \cite{schmidhuber1992learning,bengio2013representation,} and quantum state preparation \cite{1804.00633} in a single trainable pipeline that works across datasets, tasks, and model choices. We show the reconstruction results for first MNIST on 5 qubits, then CIFAR10 on 7 qubits and ImageNet on 13 qubits(Section~\ref{sec:rec_results}), demonstrating that high-dimensional data can be compressed into extremely low-dimensional quantum latent spaces while preserving recoverability. A summary of the datasets used is shown in Table~\ref{tab:datasets}.

\begin{table}[t!]
\centering
\begin{tabular}{|l|c|c|c|c|c|c|}
    \hline
    \textbf{Dataset} & \textbf{Image size} & \textbf{Channels} & \textbf{Classes} 
    & \textbf{Train} & \textbf{Validation} & \textbf{Test} \\
    \hline
    MNIST \cite{deng2012mnist}         & $28\times28$  & 1 (grayscale) & 10    & 50{,}000  & 10{,}000 & 10{,}000 \\
    CIFAR-10 \cite{CIFAR}              & $32\times32$  & 3 (RGB)       & 10    & 40{,}000  & 10{,}000 & 10{,}000 \\
    CIFAR-100 \cite{CIFAR}             & $32\times32$  & 3 (RGB)       & 100   & 40{,}000  & 10{,}000 & 10{,}000 \\
    ImageNet \cite{ILSVRC15}           & $256\times256$& 3 (RGB)       & 1{,}000 & 1{,}281{,}167 & 50{,}000 & 100{,}000 \\
    \hline
\end{tabular}
\caption{Datasets used in this work. MNIST and CIFAR train sets are split into 
training and validation subsets. ImageNet uses the standard ILSVRC splits.}
\label{tab:datasets}
\end{table}

\begin{figure}[t]
\centering
\resizebox{\textwidth}{!}{%
	
	\Qcircuit @C=0.9em @R=1.5em {
		\lstick{q_0} 
		& \multigate{2}{\text{AmplitudeEmbedding($x$)}} 
		& \gate{RZ(\alpha_{0}^{(1)})} & \gate{RY(\beta_{0}^{(1)})} & \gate{RZ(\gamma_{0}^{(1)})} & \ctrl{1} & \qw & \ctrl{2} & \qw & \qw
		& \gate{RZ(\alpha_{0}^{(2)})} & \gate{RY(\beta_{0}^{(2)})} & \gate{RZ(\gamma_{0}^{(2)})} & \ctrl{1} & \qw & \ctrl{2}& \qw  & \qw
		&  \\
		\lstick{q_1} 
		& \ghost{\text{AmplitudeEmbedding($x$)}} 
		& \gate{RZ(\alpha_{1}^{(1)})} & \gate{RY(\beta_{1}^{(1)})} & \gate{RZ(\gamma_{1}^{(1)})} & \ctrl{-1} & \ctrl{1} & \qw & \qw & \qw
		& \gate{RZ(\alpha_{1}^{(2)})} & \gate{RY(\beta_{1}^{(2)})} & \gate{RZ(\gamma_{1}^{(2)})} & \ctrl{-1} & \ctrl{1} & \qw & \qw & \qw
		& \qw \\
		\lstick{q_2} 
		& \ghost{\text{AmplitudeEmbedding($x$)}} 
		& \gate{RZ(\alpha_{2}^{(1)})} & \gate{RY(\beta_{2}^{(1)})} & \gate{RZ(\gamma_{2}^{(1)})} & \qw & \ctrl{-1} & \ctrl{-2} & \qw & \qw
		& \gate{RZ(\alpha_{2}^{(2)})} & \gate{RY(\beta_{2}^{(2)})} & \gate{RZ(\gamma_{2}^{(2)})} & \qw & \ctrl{-1} & \ctrl{-2} & \qw & \qw
		& \qw \\
	}
	
}
\caption{
	Circuit-centric Classifier with amplitude embeddings input \cite{bowles2024betterclassicalsubtleart,1804.00633}. The weight parameters $\alpha$,$\beta$, $\gamma$ are all learned. This architecture can be scaled by repeating blocks after the embedding.}
\label{fig:explicit_sel_classifier0}
\end{figure}
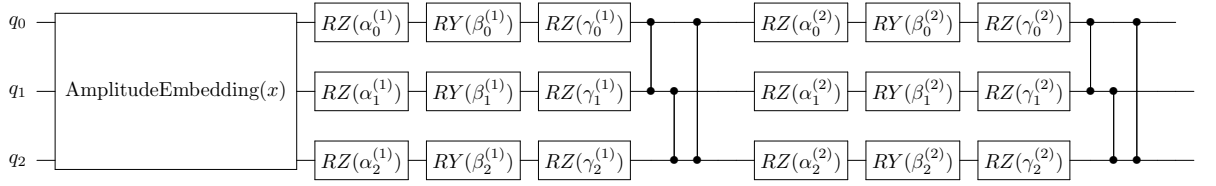

The learned embeddings are then used to train quantum classifiers (Section~\ref{sec:bench_results}). We consider two embedding strategies: an amplitude encoder, where the learned latent vector is normalized to define the quantum state amplitudes, and an ansatz encoder, where the latent variables are injected directly into the variational circuit parameters. The circuit-centric model \cite{1804.00633}, now more commonly referred to as a variational quantum classifier, is a variational quantum classifier consisting of amplitude embedding followed by repeated strongly entangled layers, as illustrated in Figure~\ref{fig:explicit_sel_classifier0}. Under identical qubit and circuit depth constraints, it achieved  over 30\%  improvement in validation accuracy on the MNIST (3 vs. 5) dataset when using the learned amplitude encoder compared to naive amplitude embedding \cite{bowles2024betterclassicalsubtleart,1804.00633,}. This result empirically validates the importance of a reconstructable latent representation; by ensuring the latent space can be mapped back to the original image, we generate a quantum state that is more structured, information-preserving, and fundamentally more trainable for machine learning tasks. Against baseline amplitude embedding, the amplitude encoder achieves an approximate exponential improvement in data recoverability, replacing exponential tomographic cost with polynomial measurement and a learned decoder (see Table~\ref{TAB:EMB}).

To address scalability, we implement the Ansatz Encoder using angle embeddings, where the learned latent variables are mapped directly to the rotation angles of single-qubit gates (Section~\ref{subsec:encoder}), forming what we previouly defined an ansatz encoder. Due to its strong empirical performance in \textcite{bowles2024betterclassicalsubtleart} we chose the data reuploading circuit \cite{P_rez_Salinas_2020}. Angle embeddings are suitable here as their implementation by single-qubit rotation gates enables the use of parameter-shift gradients \cite{Schuld_2019}. Against baseline data re-uploading, the ansatz encoder does not improve asymptotic recoverability but replaces circuit inversion with direct polynomial measurement and a learned decoder, simplifying recovery in practice on both simulated and real hardware.

In our framework, the autoencoder is trained separately and the autoencoder parameters are fixed before classification training (Section~\ref{subsec:experiments}).  This explicit training separation enables a much clearer evaluation \cite{bowles2024betterclassicalsubtleart,Mari_2020} of the quantum model: since the encoder weights are frozen, any performance gain observed during classifier training can be attributed directly to the quantum circuit rather than to continued adaptation of the classical encoder. We have not yet performed quantum classification or other learning tasks on ImageNet due to limited GPU resources. Nevertheless, the embedding results demonstrate that it is now feasible to begin obtaining results on a large, real-world dataset. 
Previously, embedding ImageNet ($256\times256\times3$, approximately 200{,}000 features) into simulable quantum circuits was not practical with standard approaches: amplitude embedding requires $O(N)$ gate depth for state preparation~\cite{1804.00633}, producing circuits too deep to simulate or execute on current hardware, while angle and data re-uploading embeddings~\cite{P_rez_Salinas_2020} require $O(N)$ qubits, far exceeding the capacity of any simulable device see Table~\ref{TAB:EMB}. Our autoencoder compresses the input to a fixed-size latent vector (13 qubits in our experiments), decoupling the embedding cost from the raw feature dimension and making simulation tractable regardless of image resolution.

We successfully executed the Ansatz Encoder on IBM quantum hardware in an inference-only mode, i.e., execution of the autoencoder using fixed trained parameters Section~\ref{subsec:experiments}). By mapping the learned latent variables to physical rotation angles, we verified that our representations remain stable and reconstructable under real-world device noise.

\section{Preliminaries and Related Work}

\subsection{Variational Autoencoders}
A Variational Autoencoder (VAE) maps input data into a latent Gaussian distribution defined by learned mean ($\mu$) and variance ($\sigma$) parameters \cite{Kingma2014}. During training, a latent vector $z$ is sampled from this distribution and passed to the decoder to reconstruct the original input.
The training objective of a VAE is to minimize the reconstruction loss, typically the mean square error, and the Kullback-Leibler (KL) divergence of the latent distribution \cite{Kingma2014}.  In our implementation for a larger dataset, we use the modern loss function from \citetitle{Rombach_2022_CVPR} \cite{Rombach_2022_CVPR}, which adds a perceptual loss \cite{PerceptualLosses2016} from \citetitle{Esser_2021_CVPR} \cite{Esser_2021_CVPR}. 

The perceptual loss compares the reconstructed images within the deep-layer representations of a pre-trained neural network, typically VGG16 \cite{simonyan2015VGG}. This prioritizes the preservation of semantic structures---the high-level features and spatial relationships within the image. To implement this, we employ the Learned Perceptual Image Patch Similarity (LPIPS) metric \cite{PerceptualLosses2016}, enabling the use of high-fidelity, modern VAE architectures like those in Latent Diffusion Models \cite{Rombach_2022_CVPR}. 
Additionally, an adversarial loss is employed, where a discriminator is trained to distinguish between reconstructed and original images. This forces the VAE to ``fool'' the discriminator \cite{Isola_2017_CVPR}, thereby reducing blurriness and preserving photorealistic textures.

\subsection{Quantum Classifier}
Quantum classifier is a machine learning model that uses quantum computing resources to label input data. In this work, we consider a Variational Quantum Classifier (VQC) framework, which utilizes a Parameterized Quantum Circuit (PQC) as the core model.

\begin{figure}[ht!]
	\centering
	\begin{subfigure}{0.90\textwidth}
		\centering
		\begin{tabular}{cc}
			\amplitudeembedding & \stronglyentangledlayercircuit  \\
			Amplitude embedding of input $x$. & A single strongly entangled layer
		\end{tabular}
	\end{subfigure}
	\begin{subfigure}{\textwidth}
		\centering
		\resizebox{\textwidth}{!}{
			\(
			\Qcircuit @C=0.9em @R=1.5em {
				\lstick{q_0} 
				& \multigate{2}{\text{AmplitudeEmbedding($x$)}} 
				& \multigate{2}{\text{Strongly Entangled Layer~$(\theta)_1$}} 
				&\qw & \cdots &
				& \multigate{2}{\text{Strongly Entangled Layer~$(\theta)_i$}} 
				&\qw & \cdots &
				& \multigate{2}{\text{Strongly Entangled Layer~$(\theta)_n$}} 
				& \meter \\
				\lstick{q_1} 
				& \ghost{\text{AmplitudeEmbedding($x$)}} 
				& \ghost{\text{Strongly Entangled Layer~$(\theta)_1$}} 
				&\qw & \cdots &
				& \ghost{\text{Strongly Entangled Layer~$(\theta)_i$}} 
				&\qw & \cdots &
				& \ghost{\text{Strongly Entangled Layer~$(\theta)_n$}} 
				& \meter \\
				\lstick{q_2} 
				& \ghost{\text{AmplitudeEmbedding($x$)}} 
				& \ghost{\text{Strongly Entangled Layer~$(\theta)_1$}} 
				&\qw & \cdots  &
				& \ghost{\text{Strongly Entangled Layer~$(\theta)_i$}} 
				&\qw & \cdots &
				& \ghost{\text{Strongly Entangled Layer~$(\theta)_n$}} 
				& \meter \\
			}
			\)
		}
	\end{subfigure}
	\caption{Quantum Classifier Circuit \cite{10.5555/3309066}: amplitude embedding followed by repeated strongly entangled layers. This model is scaled by the number of repeated strongly entangled layers.}
	\label{fig:amplitude_classifier}
\end{figure}
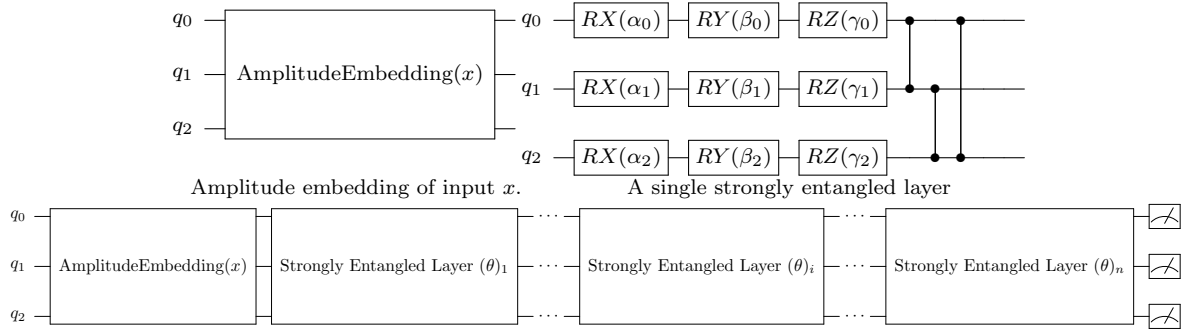

Classical data is first embedded into a quantum state in Hilbert space using encoding strategies such as amplitude or angle embedding. The resulting state is then processed by a parameterized quantum circuit, often referred to as the ansatz, which defines the circuit architecture and trainable operations. The choice of ansatz is a critical hyperparameter, as it directly impacts the expressivity and trainability of the model. 
Finally, measurements are performed on the output quantum state, and the expectation values are mapped to class predictions \cite{1804.00633}. An example of a quantum classifier using amplitude embedding is shown in Fig.~\ref{fig:amplitude_classifier}.

\subsubsection{Quantum Gradient} 

Quantum gradients are estimated using two techniques. Finite difference is a traditional numerical method to find the derivative. The second is the parameter-shift rules, which provides an exact method to compute gradients for certain parameterized quantum gates, such as single-qubit rotation gates \cite{Schuld_2019,1804.00633, 10.5555/3309066}.

For example, consider the following cost function defined as the expectation value of the measurement \begin{equation}
f(\theta) = \bra{0} U^\dagger\!(\theta)\hat{O}U(\theta)\ket{0}.\end{equation} For a single qubit rotation gate $U(\theta)$, the derivative with respect to $\theta$ is computed by 
\begin{equation}
\frac{\partial f(\theta)}{\partial \theta}
=
\frac{1}{2}
\left[
f\left(\theta + \frac{\pi}{2}\right)
-
f\left(\theta - \frac{\pi}{2}\right)
\right].
\end{equation}
Thus, the gradient is obtained by executing the circuit twice with both $\theta + \frac{\pi}{2} $ and $\theta - \frac{\pi}{2} $. This rule applies to more gates than singlequbit Pauli rotation \cite{ 
1804.00633,Schuld_2019,, 10.5555/3309066}, and finding gates where the parameter shift rule applies, or other analytic gradient methods, is an active area of research~\cite{wierichs2022general, banchi2021stochastic, kyriienko2021generalized, wiersema2024sun}.

\subsubsection{Data Reuploding Classifier}

The Data Reuploding Classifier from 
\citetitle{P_rez_Salinas_2020}
\cite{P_rez_Salinas_2020} uses a specific topology where the input is embedded throughout the circuit, not just at the initial phase. For each layer of the circuit, part of the input is fed into angle embeddings (rotation gates). 

\subsection{Embeddings}

Amplitude embeddings encode data on the amplitude of a quantum state. First, the data are flattened and normalized, and each value is encoded into a $\alpha_i$ on the quantum state: $\psi= \sum \alpha_i\ket{i}$. For inputs that are not exactly powers of 2, constants can be added to pad the state to a power of 2. In contrast, angle embeddings map a feature $x_i$ into one or more rotation gates $R_\alpha(x_i)$. Typically, single qubit rotation gates are used, i.e $\alpha \in \{X, Y, Z\} $.

\section{Variational Quantum Autoencoders for Tailor-Made Embeddings }

\begin{figure}[h]
	\centering
	\includegraphics[scale=0.25]{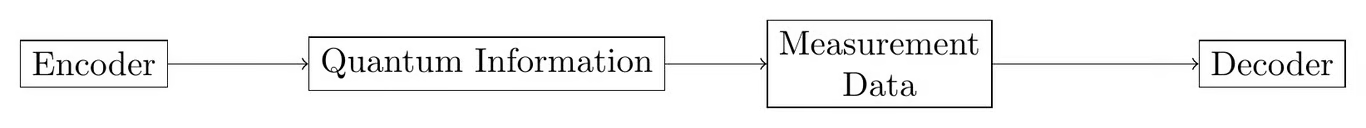}\\
	\caption{Variational Quantum Latent Autoencoders. The encoder embeds the data into quantum information. The decode use measurement data  extracted from this quantum information to reconstruct the original classical data.}
	\label{fig:qvaegeneral}
\end{figure}

Our approach consists of a variational autoencoder with a quantum latent space, as illustrated in Figure~\ref{fig:qvaegeneral}. The role of the autoencoder is to compress the input data into a low-dimensional representation that can be processed by a quantum model. Depending on the chosen encoding, this latent representation may correspond to the amplitudes of a quantum state, the parameters of a variational quantum circuit, a Hamiltonian, or another quantum object.

The framework is designed to be independent of the specific quantum encoding. Instead, lightweight projection layers transform the bottleneck representation into the format required by the target quantum system, matching its dimensionality, normalization, and parameter ranges. As a result, the same classical autoencoder architecture can be combined with a wide variety of quantum embeddings without modifying the overall framework.

\subsection{Encoder}
\label{subsec:encoder}

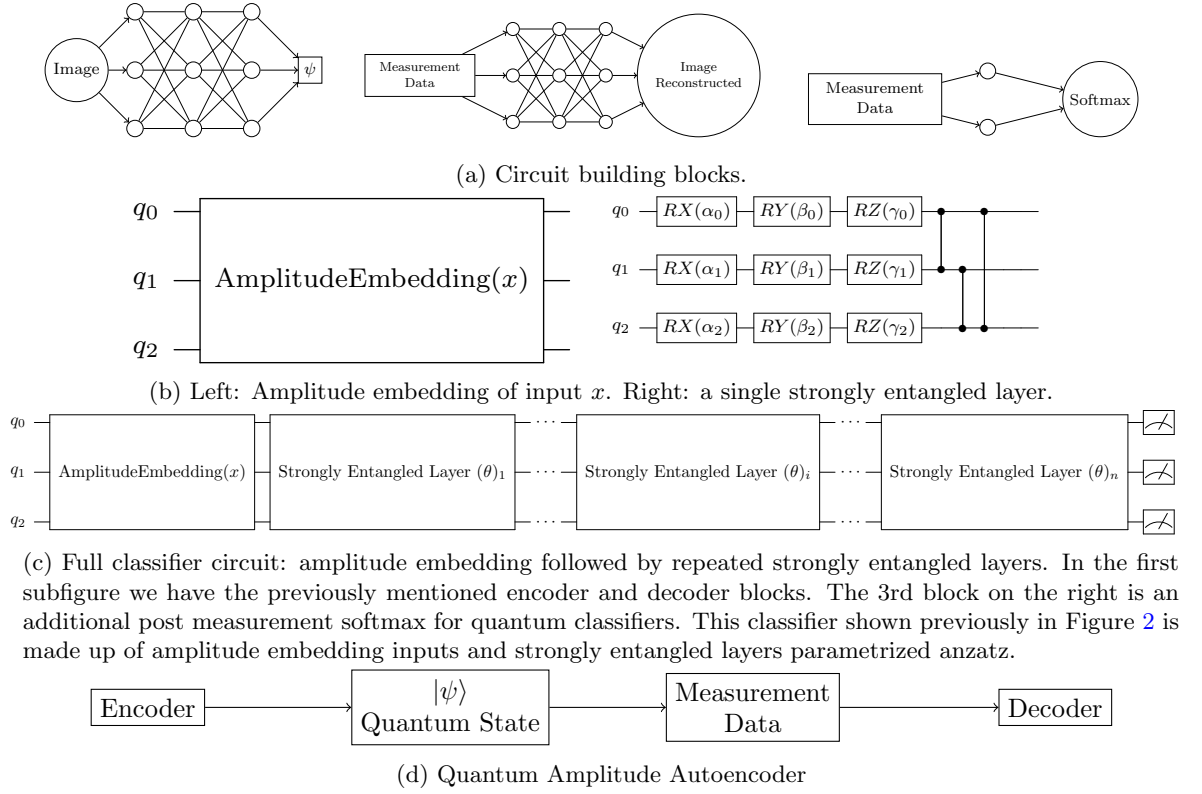
\begin{figure}[t!]  
\begin{subfigure}{\textwidth}
	\begin{tabular}{ccc}
		\resizebox{0.25\textwidth}{!}{\encoder{$\psi$}}&  \resizebox{0.35\textwidth}{!}{\decoder}& \resizebox{0.30\textwidth}{!}{ \postprocessinglayer[\tiny{\twoline{Measurement}{Data}}]}
	\end{tabular}
	
	\caption{Circuit building blocks.}
\end{subfigure}

\begin{subfigure}{\textwidth}
	\centering
	\begin{tabular}{ccc}
		\resizebox{0.35\textwidth}{!}{\amplitudeembedding} &\hfill 
		&     \resizebox{0.35\textwidth}{!}{\stronglyentangledlayercircuit} 
	\end{tabular}
	\caption{Left: Amplitude embedding of input $x$. Right: a single strongly entangled layer.}
\end{subfigure}


\begin{subfigure}{\textwidth}
	\centering
	\resizebox{\textwidth}{!}{
		\amplitudeclassifier
	} 
	\caption{Full classifier circuit: amplitude embedding followed by repeated strongly entangled layers. In the first subfigure we have the previously mentioned encoder and decoder blocks. The 3rd block on the right is an additional post measurement softmax for quantum classifiers. This classifier shown previously in Figure \ref{fig:amplitude_classifier} is made up of amplitude embedding inputs and strongly entangled layers parametrized anzatz. }
	\label{fig:full_classifier}
\end{subfigure}

\begin{subfigure}{\textwidth}
	\centering
	\begin{tikzpicture}[grow=right, node distance=3cm and 4cm, every node/.style={align=center}]
		\node[rectangle,draw] (IMGin) at (0,0) {Encoder};
		\node[rectangle,draw] (QState) at (4,0) {$\ket{\psi}$\\Quantum State};
		\node[rectangle,draw] (Measure) at (8,0) {Measurement\\Data};
		\node[rectangle,draw] (Decoder) at (12,0) {Decoder};
		
		\draw[->] (IMGin) -- (QState);
		\draw[->] (QState) -- (Measure);
		\draw[->] (Measure) -- (Decoder);
	\end{tikzpicture}
	\caption{Quantum Amplitude Autoencoder}
	\label{fig:subfig_a}
\end{subfigure}
\caption{
Encoder and decoder architecture of the amplitude autoencoder. The encoder maps classical data into a quantum state, while the decoder reconstructs the original images from measurement outcomes obtained from that quantum state.}
\label{fig:amqvaesimple}
\end{figure}

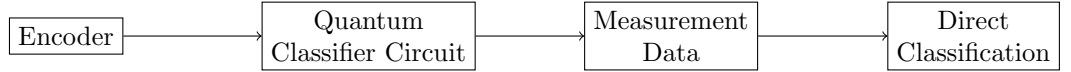
\begin{figure}[t]
\centering
\begin{tikzpicture}[grow=right, node distance=3cm and 4cm, every node/.style={align=center}]
	\node[rectangle,draw] (IMGin) at (0,0) {Encoder};
	\node[rectangle,draw] (QClass) at (4,0) {Quantum\\Classifier Circuit};
	\node[rectangle,draw] (Measure) at (8,0) {Measurement\\Data};
	\node[rectangle,draw] (QClassOut) at (12,0) {Direct\\Classification};
	
	\draw[->] (IMGin) -- (QClass);
	\draw[->] (QClass) -- (Measure);
	\draw[->] (Measure) -- (QClassOut);
\end{tikzpicture}
\caption{The encoder feed the input to the quantum classifiers. Measurement on select registers gives the classification results.  }
\label{fig:anqvaesimple}
\end{figure}

Starting from a classical variational encoder, which is a neural network that takes the binary encoding of an image and generates and encodes it into the data being processed. The quantum autoencoder differs in that while it is still a classical neural network that does the encoding, now, the output is either a set of numbers that define amplitudes of a quantum state, or a set of parameters that parametrize a quantum circuit. We change the output from classical space to a quantum state or a group of quantum states. 
We modify the final layers of classical encoder to output the physical instructions required to initialize a quantum state from classical input data. We define two distinct architectural modes based on the chosen embedding strategy:

\begin{itemize}
    \item \textbf{Amplitude Encoder (State Description):} In this configuration, the classical neural network outputs a vector of complex amplitudes $\psi$. These are used for state preparation, allowing the input data to be embedded directly into the probability amplitudes of the quantum state. This method leverages the exponential capacity of the quantum Hilbert space. Figure~\ref{fig:amqvaesimple} shows a high level view of the model. 
    
    \item \textbf{Ansatz Encoder (Input Parameters):} Here, the encoder outputs a set of rotation angles $\theta$ that serve as the \textit{input parameters} for the quantum circuit. This approach ensures the framework is scalable for a larger number of qubits, as it avoids the exponential memory requirements associated with classical descriptions of full quantum states. Figure~\ref{fig:anqvaesimple} shows a high level view of the model. 
\end{itemize}

We choose a VAE over a standard autoencoder for reasons motivated by quantum state preparation requirements. The KL regularization is designed to encourage a smooth latent manifold \cite{Kingma2014}, reducing the risk of discontinuities that could produce unstable quantum states from nearby inputs, a property standard autoencoders do not explicitly optimize for. Additionally, we adopt the autoencoder-KL variant \cite{Rombach_2022_CVPR} which incorporates perceptual loss \cite{PerceptualLosses2016}, with the expectation that semantic structure relevant to downstream classification is better preserved than with pixel-level MSE alone. This expectation is indirectly supported by the finding of \cite{lu2025fidelitypreserving} that SSIM, a measure of structural fidelity, correlates more strongly with downstream QNN classification performance than pixel-level metrics such as PSNR or MSE, suggesting that preserving semantic structure during encoding is beneficial for quantum learning tasks, though a direct causal link remains to be established. We note that this supporting evidence from \cite{lu2025fidelitypreserving} emerged after the present work was conducted. Our choice of VAE with perceptual loss was made on architectural grounds, and the SSIM finding provides independent, subsequent corroboration. We leave a systematic ablation of encoder architecture choices, including standard AE, VAE, and VQ-VAE, to future work.

\begin{figure}[t]
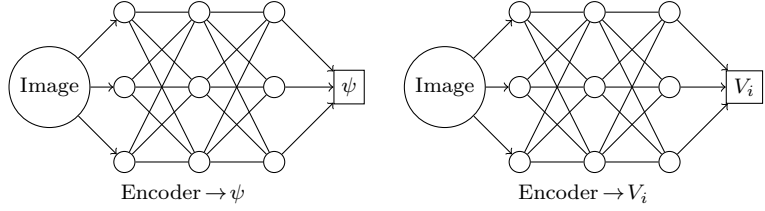

	\centering
	\resizebox{0.70\textwidth}{!}{
		\begin{tabular}{cc}
			\encoder{$\psi$}&  \encoder{$V_i$} \\
			Encoder$\,\rightarrow \!\!\,\psi$ & Encoder$\,\rightarrow \!\!\,V_i$
	\end{tabular}}
	\caption{Encoder choice. On the left side we show an amplitude encoder where a neural network generate the state description $\psi$. On the right side we show an anzatz encoder where the neural networks generated a sets of quantum circuit input $V_i$ }
	\label{fig:encoderchoice}
\end{figure}

For the Amplitude Encoder We can modify a softmax layer to output a state amplitude by solving this equation $\alpha_k = \frac{e^{\mathrm{i}x_k}}{\sqrt{\sum_j |e^{\mathrm{i}x_j}|^2}}$.
As illustrated in Figure \ref{fig:encoderchoice}, both architectures enable us to``plug'' high-performance classical models, such as CNNs with attention mechanisms \cite{Esser_2021_CVPR, Rombach_2022_CVPR}, directly into a quantum latent space by simply adapting the final output layer to meet the numerical constraints of the target quantum system parameters.

\subsection{Decoder}

\begin{figure}[t]
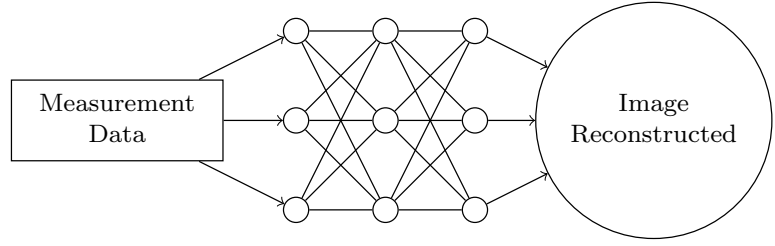

	\centering
	\resizebox{0.70\textwidth}{!}{
		\begin{tabular}{c}
			\decoder
	\end{tabular}}
	\caption{The decoder is a neural network that takes measurement data and reconstructs the original data, in this case an image. }
	\label{fig:Decoder}
\end{figure}

The decoder provides the quantum-to-classical interface by reconstructing the original input from measurement outcomes obtained from the quantum latent state. It is implemented as a standard neural network whose input layer is adapted to process quantum measurement data rather than a classical latent vector. Depending on the chosen measurement strategy, the decoder may receive quantities such as probability distributions or expectation values. Figure~\ref{fig:Decoder} illustrates a simple multilayer perceptron decoder, although in practice more expressive architectures such as CNNs~\cite{simonyan2015VGG,Isola_2017_CVPR} or attention-based models~\cite{Esser_2021_CVPR,Rombach_2022_CVPR} can be used.

\subsection{Measurement Strategies}

The measurement strategy determines which classical information is extracted from the quantum latent state before being passed to the decoder. In principle, complete quantum state tomography could recover the full wavefunction, but its cost scales exponentially with the number of qubits and is therefore impractical for large systems. Consequently, we restrict ourselves to experimentally realistic measurement schemes requiring only polynomial resources, even in simulation where the full quantum state is available.

In this work, we consider two measurement strategies. First, we use computational-basis probability distributions, which have exponentially many outcomes but can be estimated with polynomially many samples using shadow tomography~\cite{aaronson2018shadowtomographyquantumstates,Huang_2020}. Second, we use single-qubit expectation values, a common and hardware-efficient choice in quantum machine learning that can be estimated accurately with a modest number of shots (e.g., \(10^3\)–\(10^4\)). Our framework is sufficiently general to accommodate other measurement protocols, including shadow-based methods, provided they can be incorporated into gradient-based optimization.

\begin{figure}[t]
	\begin{subfigure}[b]{\textwidth}
		\centering
		\resizebox{0.85\textwidth}{!}{
			\begin{tabular}{ccc}
				\reuploadlayer &\vspace{1em}& \finalmeasurementlayer\\ 
				Reupload Layer &\vspace{1em} & Final Mesurement Layer
		\end{tabular}}
		
	\end{subfigure}
	
	\begin{subfigure}[b]{\textwidth}
		\centering
		\resizebox{0.75\textwidth}{!}{
			\Qcircuit @C=1em @R=1.5em {
				\lstick{q_0} 
				& \multigate{2}{\twoline{Reupload Layer}{$j=0$}} 
				&\qw& \cdots  &
				& \multigate{2}{\twoline{Reupload Layer}{$j=n-1$}} 
				& \multigate{2}{\twoline{Final Measurement}{Layer}} 
				\\
				\lstick{q_1} 
				& \ghost{\twoline{Reupload Layer}{$j=0$}} 
				& \qw & \cdots &
				& \ghost{\twoline{Reupload Layer}{$j=n-1$}} 
				& \ghost{\twoline{Final Measurement}{Layer}} 
				\\
				\lstick{q_2} 
				& \ghost{\twoline{Reupload Layer}{$j=0$}} 
				&\qw & \cdots  &
				& \ghost{\twoline{Reupload Layer}{$j=n-1$}} 
				& \ghost{\twoline{Final Measurement}{Layer}} 
			}
		}
		\caption{Data Reuploading Circuit}

	\end{subfigure}
	\centering
	\begin{subfigure}{\textwidth}  
		\resizebox{0.85\textwidth}{!}{
			\begin{tabular}{cc}
				\encoder{$V_i$}& \decoder \\
				Encoder$\,\rightarrow \!\!\,V_i$ & Decoder
		\end{tabular}}
	\end{subfigure}
	
	\begin{subfigure}{\textwidth}
		\begin{tikzpicture}[grow=right]
			\node[rectangle,draw,] (Enc) at (0, 0)  {Encoder$\,\rightarrow \!\!\,V_i$};
			
			Data reuploading Quantum classifier [15]
			
			\node[rectangle, draw] (Cr) at (6, 0) {
				Data Reuploading Circuit
			};
			
			\node[rectangle,draw,] (Dec) at (12, 0)  {Decoder};
			
			\path [->] (Enc) edge node[near start,above] {} (Cr);
			
			\path [->] (Cr) edge node[near end,above] {} (Dec);

		\end{tikzpicture}	
\end{subfigure}

\caption{Variational Quantum Autoencoder building blocks. The Data reuploading classifier layers are at the top. The classical autoencoder architecture encodes data within the circuit input parameters of the Data reuploading Quantum classifier \cite{P_rez_Salinas_2020}.}
\label{fig:anzatzvae}
\end{figure}

\subsection{Quantum VAE}

Figure~\ref{fig:anzatzvae} illustrates the quantum VAE using a data re-uploading ansatz. In this architecture, the latent representation produced by the encoder is mapped directly to the rotation angles $(x_i^j,y_i^j,z_i^j)$ of the gates in each data re-uploading layer. Consequently, each point in the latent space specifies a particular parameterized quantum circuit, and the latent space can be interpreted as the set of circuits realizable within the chosen ansatz.

The trainable circuit parameters $w_i^j$, which are later used for downstream classification, can either be frozen or optimized during the autoencoder training stage. Similar to unsupervised pre-training in classical neural networks, this procedure provides an informed initialization that can improve subsequent optimization and learning performance~\cite{UnsupervisedPretraining}.

\subsection{Training}

The training procedure largely follows that of classical VAEs. For the MNIST dataset, we use the mean squared error (MSE) as the reconstruction loss. For the more challenging CIFAR-10, CIFAR-100, and ImageNet datasets, we employ the LPIPS perceptual loss together with adversarial training, as described previously. In all cases, the VAE is regularized using the standard KL-divergence term on the latent distribution.

For amplitude encodings, the encoding operation can be implemented entirely with classical linear algebra, avoiding quantum circuit simulation during training and making optimization significantly more efficient. In contrast, when the latent representation specifies the parameters of a variational quantum circuit, quantum simulation is unavoidable.

If trained on quantum hardware, the gate-parameter encoder requires gradient estimation using the parameter-shift rule. In classical simulation, however, more efficient techniques such as backpropagation or the adjoint differentiation method can be used, offering substantial computational advantages. By keeping as much of the training pipeline as possible in the classical domain, we reduce the impact of common quantum machine learning challenges, including barren plateaus and the high cost of gradient evaluation associated with parameter-shift or finite-difference methods.

\subsubsection{Pre-Training}
Unsupervised pre-training is a classical technique in which a model's weights are first optimized on an auxiliary task—such as reconstruction—before being used to initialize training on the target task, often improving convergence and final performance \cite{UnsupervisedPretraining, bengio2007greedy}.
We apply the same principle to the ansatz encoder: the trainable circuit parameters $w_i^j$ of the data re-uploading circuit are updated jointly with the rest of the autoencoder during training (Figure~\ref{fig:anzatzvae}), rather than left at their random initial values. This provides the quantum circuit with informed initial parameters before classification training, mirroring the role of pre-training in classical networks. A more comprehensive 
investigation comparing pre-training against standard random 
initialization is planned for future work

\subsubsection{Reusing Classical Autoencoder}

With a projection layer as required, we can use the modern Hugging Face's autoencoder KL \cite{Kingma2014,Rombach_2022_CVPR} model. The projection layers are linear layers that connect the latent space to the quantum space (states or gate parameters) with normalization as required. This allows us to use a pretrained model and access advanced loss functions like LPIPS \cite{PerceptualLosses2016,Rombach_2022_CVPR,zhang2018unreasonableeffectivenessdeepfeatures}. In our experiments, we train our own model using the LPIPS Loss function. Doing transfer learning with pretrained classical models is a task for future work. Reusing and finetuning pretrained models is standard in modern production machine learning. Training a large VAE is a complex  learning task with significant hyperparameter sensitivity~\cite{rombach2021highresolution}. With computing power already being one of our major limitations, starting from an already trained classical model might lead to better results with less computational effort,
as the model would not need to relearn general image representations from scratch.

\subsection{Experiments}
\label{subsec:experiments}

Two distinct autoencoder architectures were employed depending on dataset 
scale. For MNIST, a custom VAE was used in which the encoder is a 
two-layer neural network ($784 \rightarrow 512 \rightarrow 256$ with 
LayerNorm and ReLU activations) that compresses each input image into a 
compact latent representation. Rather than producing a single fixed 
vector, the encoder outputs the parameters of a probability distribution 
following the standard VAE approach \cite{Kingma2014} from which 
a latent vector is sampled during training. This sampling is applied 
independently to both the real and imaginary parts of the latent vector, 
which are then normalized as described in Section~\ref{subsec:encoder} to 
produce a valid quantum state. The decoder is a neural network that takes 
the quantum measurement outputs and reconstructs the original image from 
them. The training loss combines MSE reconstruction loss, a KL divergence 
term, and a cycle consistency term that measures quantum state fidelity 
between the encoded input state and the re-encoded reconstruction, each 
weighted at $10^{-4}$.
For CIFAR-10 and ImageNet, we adopt the AutoencoderKL architecture from 
Latent Diffusion Models \cite{Rombach_2022_CVPR} with LPIPS perceptual 
loss \cite{PerceptualLosses2016} and KL divergence weighted at $10^{-8}$. 
The cycle consistency term used in the MNIST architecture was omitted 
here, as the richer LPIPS and adversarial objectives provided sufficient 
regularization and the additional fidelity term introduced competing 
gradient signals. Reintegrating the cycle consistency loss with properly 
tuned weighting across all datasets remains a direction for future work.

All models were trained using the Adam optimizer with gradient clipping 
at a maximum value of $1.0$. Learning rates were $10^{-3}$ for the MNIST 
autoencoder and $10^{-4}$ for CIFAR-10 and ImageNet autoencoders. 
Classification models used learning rates of $10^{-4}$ for VAE-based 
models and $10^{-3}$ for circuit-centric classifiers. The training was for up to 1{,}000 epochs with a batch size of 128, except for the ansatz encoder experiments, which used a batch size of 32, and the ImageNet autoencoder, which was trained for 5 epochs due to the high computational cost of training at this scale. Training for ImageNet was stopped at this point as the reconstruction results were already visually satisfactory, despite significant room for further improvement with continued training. The model checkpoint with the lowest validation perceptual loss was selected for evaluation in all other cases.

Data splits were as follows: MNIST used 50,000 training / 10,000 
validation / 10,000 test samples; the binary 3-vs-5 classification subset 
used a 90/10 train/validation split from the filtered set; CIFAR-10 used 
40,000 / 10,000 / 10,000. Gaussian noise augmentation with standard 
deviation $0.1$ was applied as a denoising objective 
\cite{vincent2008extracting} after a warmup period of 60 epochs for MNIST 
and 6 epochs for CIFAR-10.

\begin{figure}[t]
\centering
\newlength{\ibmimgw}
\setlength{\ibmimgw}{0.11\textwidth}
\setlength{\tabcolsep}{4pt}
\begin{tabular}{cccc}
    \multicolumn{4}{c}{\textbf{IBM QPU Inference }} \\[4pt]
    \includegraphics[width=\ibmimgw]{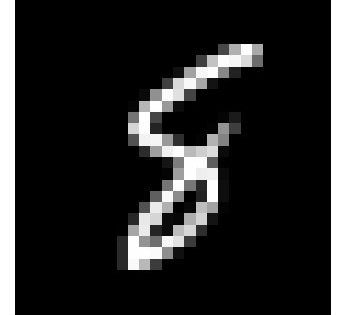} &
    \includegraphics[width=\ibmimgw]{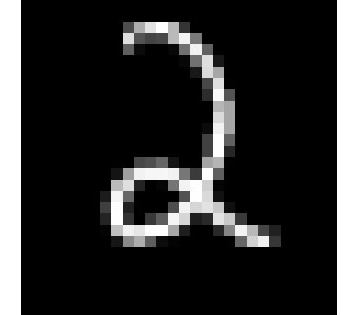} &
    \includegraphics[width=\ibmimgw]{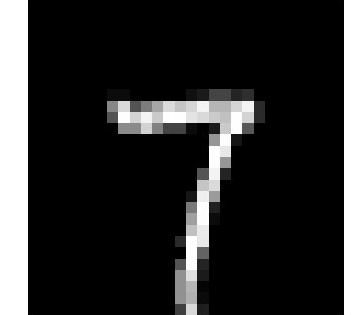} &
    \includegraphics[width=\ibmimgw]{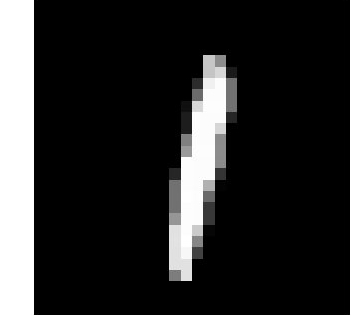} \\[4pt]
    \includegraphics[width=\ibmimgw]{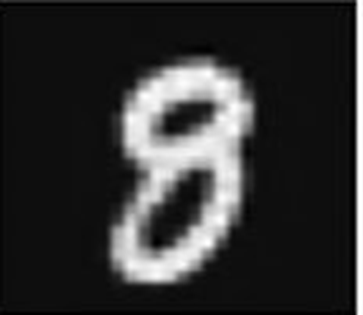} &
    \includegraphics[width=\ibmimgw]{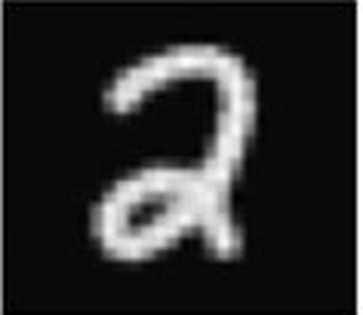} &
    \includegraphics[width=\ibmimgw]{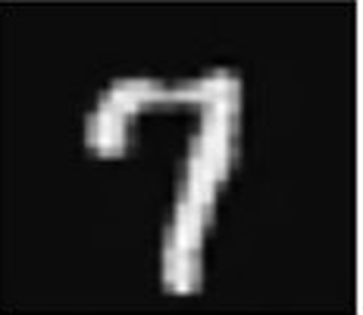} &
    \includegraphics[width=\ibmimgw]{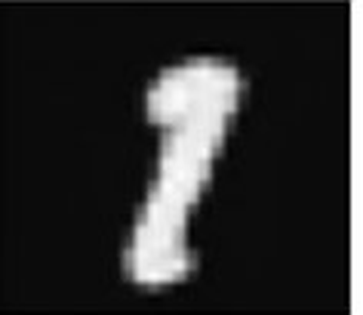} \\
\end{tabular}
\caption{IBM Yonsei quantum hardware inference results. Original MNIST test samples (top) and reconstructions decoded from quantum measurements (bottom). The ansatz encoder was executed in inference-only mode with fixed trained parameters. }
\label{fig:ibm_results}
\end{figure}

The ansatz encoder circuit used for MNIST pretraining used 10 qubits and 
10 data reuploading layers \cite{P_rez_Salinas_2020}, with alternating 
double and double-odd CZ entanglement patterns between layers. The 
CIFAR-10 ansatz encoder also used 10 qubits under the same topology. All 
quantum simulations used PennyLane's \texttt{default.qubit} simulator 
with backpropagation-compatible gradients \cite{Schuld_2019}. The 
amplitude embedding baseline was stopped at approximately 250 epochs as 
the validation loss showed no improvement throughout training, as visible 
in Figure~\ref{fig:vallosscifar}; all other models completed the 
full 1{,}000 epochs under identical training conditions.
The MNIST ansatz encoder was subsequently executed on IBM quantum hardware 
in inference-only mode using fixed trained parameters. The execution was 
performed as part of a proprietary collaboration and full hardware 
configuration details are not available for publication; qualitative 
reconstruction results confirming stability under device noise are shown 
in Figure~\ref{fig:ibm_results}.

Quantum classification experiments were performed on MNIST, CIFAR-10, 
and CIFAR-100. For CIFAR-100, the amplitude encoder trained on CIFAR-10 
was reused without retraining, with the quantum classifier trained using 
the Adam optimizer at a learning rate of $10^{-5}$. This cross-dataset 
transfer tests the generalization of the learned embedding to a 
higher-complexity 100-class problem under the same encoder. Classification 
on ImageNet was not performed due to the prohibitive computational cost 
of quantum circuit simulation at that scale. 

\subsection{Comparison}

Quantum embeddings differ significantly in their circuit depth, qubit requirements, and measurement complexity. The proposed autoencoder framework is designed to be agnostic to the specific embedding and can be adapted to a wide range of encoding strategies, circuit architectures, and measurement protocols. Likewise, the amount and type of measurement data used for reconstruction can be tailored to the target application.

For this reason, we refer to our approach as providing \emph{tailor-made embeddings}: the learned representation can be customized to the available quantum resources and the characteristics of the underlying data. Although our experiments focus on image datasets, the framework is not restricted to this domain and could, in principle, be applied to other forms of classical or physically motivated data. Whether the learned embeddings preserve task-specific or physically meaningful properties, however, is ultimately an empirical question and may require alternative autoencoder architectures or training objectives.

\begin{figure}[t]
\centering
\begin{tikzpicture}[grow=right, node distance=3cm and 4cm, every node/.style={align=center}]
	\node[rectangle,draw] (IMGin) at (0,0) {Encoder};
	\node[rectangle,draw] (QClass) at (4,0) {Quantum\\Classifier Circuit};
	\node[rectangle,draw] (Measure) at (8,0) {Measurement\\Data};
	\node[rectangle,draw] (Softmax) at (12,0) {Softmax};
	
	\draw[->] (IMGin) -- (QClass);
	\draw[->] (QClass) -- (Measure);
	\draw[->] (Measure) -- (Softmax);
\end{tikzpicture}
\caption{The encoder feed the input to the quantum classfier. The measurement data is process by a classical softmax layer.} 
\label{fig:subfig_c}
\end{figure}
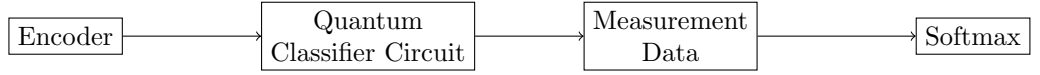

\begin{table}[t]
\centering
\begin{tabular}{|l|l|c|c|c|c|c|}
\hline
\textbf{Train Data} & \textbf{Test Data} & \textbf{Resolution} & \textbf{PSNR (dB) $\uparrow$} & \textbf{SSIM $\uparrow$} & \textbf{LPIPS $\downarrow$} & \textbf{FID $\downarrow$} \\
\hline
CIFAR-10 & CIFAR-10 & $32\times32$ & $14.06 \pm 2.35$ & $0.287 \pm 0.079$ & $0.314 \pm 0.053$ & $133.7$ \\
CIFAR-10 & CIFAR-100 & $32\times32$ & $13.49 \pm 2.60$ & $0.276 \pm 0.084$ & $0.353 \pm 0.061$ & --- \\
ImageNet & CIFAR-10 & $256\times256$ & $35.21 \pm 5.48$ & $0.969 \pm 0.052$ & $0.0078 \pm 0.0049$ & $0.61$ \\
\hline
\end{tabular}
\caption{Reconstruction metrics for amplitude encoder. $\pm$ values denote standard deviation across images in the test set. The ImageNet to CIFAR-10 transfer learning result should be interpreted with caution as CIFAR-10 images were upsampled from $32\times32$ to $256\times256$ prior to encoding, which substantially reduces reconstruction difficulty compared to the native resolution experiments as it increase the size of the model and the quantum latent space. This explains the near Zero FID. }
\label{tab:reconstruction_metrics}
\end{table}

\begin{figure}[h!]
\centering       
\begin{tabular}{cccc}
	\includegraphics[scale=0.33]{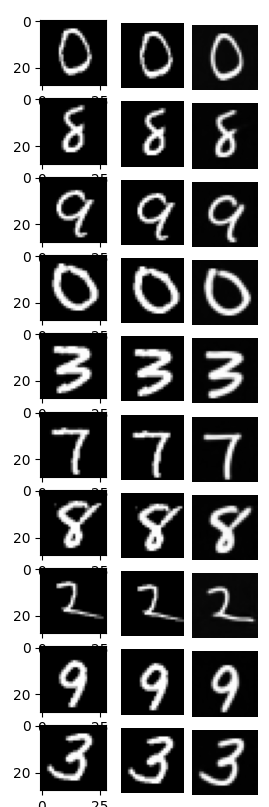} & 
	\includegraphics[scale=0.33]{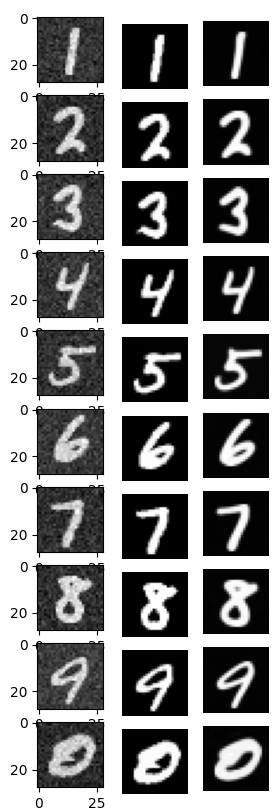}  
	\includegraphics[scale=0.33]{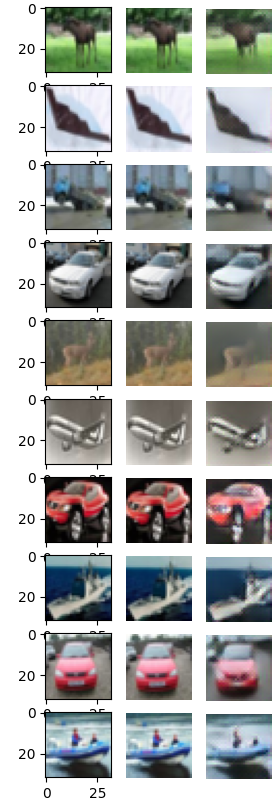} & 
	\includegraphics[scale=0.33]{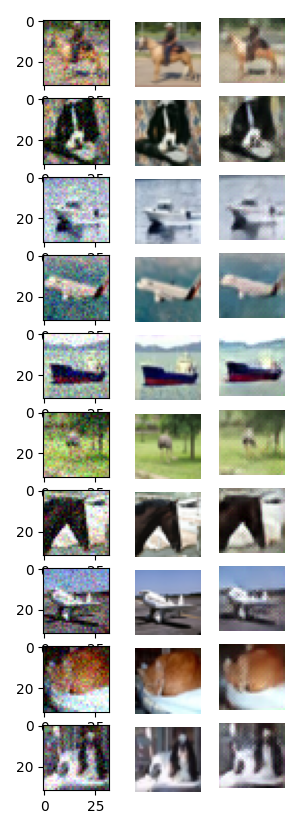}  
\end{tabular}
\caption{Autoencoder results on the MNIST and CIFAR-10 datasets. The left side corresponds to the validation set and the right side to the test set. The first column shows the
input images, which may be noisy, the second column the original images, and the third column the reconstructed images. }
\label{fig:VAECIFARMNIST}
\end{figure}

\section{Reconstruction Results}
\label{sec:rec_results}
We evaluate reconstruction quality using four complementary metrics. 
Peak Signal-to-Noise Ratio (PSNR) measures pixel-level fidelity in decibels, 
where higher values indicate closer reconstruction to the original image. 
Structural Similarity Index (SSIM)~\cite{wang2004ssim} measures the preservation 
of structural information, luminance, and contrast, ranging from 0 to 1 where 
1 indicates perfect reconstruction. Learned Perceptual Image Patch Similarity 
(LPIPS)~\cite{zhang2018unreasonableeffectivenessdeepfeatures} measures perceptual 
similarity using features from a pretrained VGG network~\cite{simonyan2015VGG},
where lower values indicate more perceptually similar reconstructions. 
Fr\'{e}chet Inception Distance (FID)~\cite{heusel2017fid} measures the 
distributional similarity between reconstructed and original images, where 
lower values are better. The training loss combines LPIPS and KL 
divergence~\cite{Kingma2014}. Full adversarial training following~\cite{Rombach_2022_CVPR} 
is left for future work.

\begin{figure}[t]
\newlength{\imgw}
\setlength{\imgw}{0.16\textwidth}
\setlength{\tabcolsep}{1pt}
\centering 
	\begin{tabular}{cccccc}
	  \textbf{Original} & \textbf{Reconstruction}& \textbf{Original} & \textbf{Reconstruction}&  \textbf{Original} & \textbf{Reconstruction}\\	\includegraphics[width=\imgw]{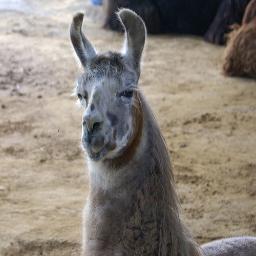} & 
		\includegraphics[width=\imgw]{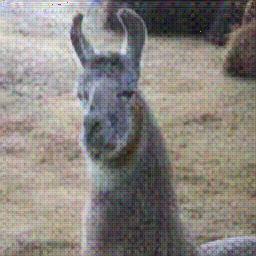} & 
		\includegraphics[width=\imgw]{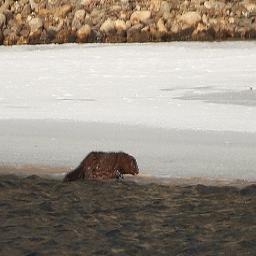} & 
		\includegraphics[width=\imgw]{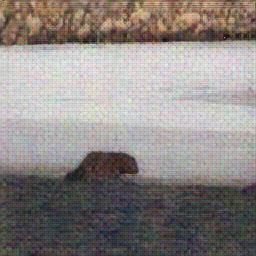} & 
		\includegraphics[width=\imgw]{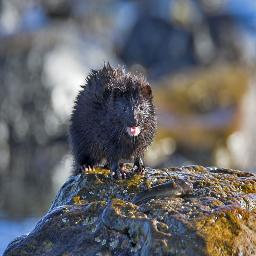} & 
		\includegraphics[width=\imgw]{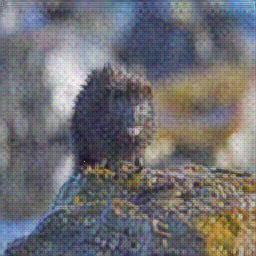} \\
		\includegraphics[width=\imgw]{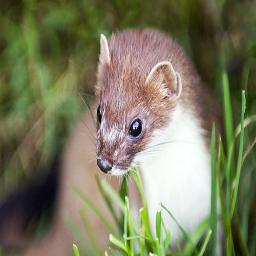} & 
		\includegraphics[width=\imgw]{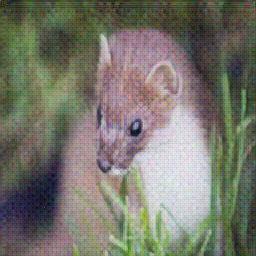} & 
		\includegraphics[width=\imgw]{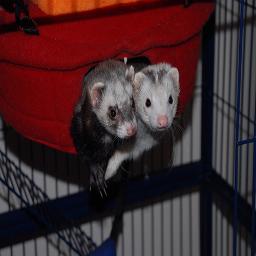} & 
		\includegraphics[width=\imgw]{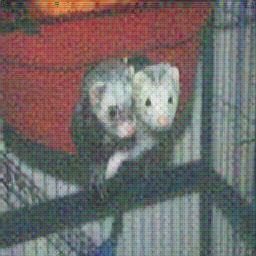} & 
		\includegraphics[width=\imgw]{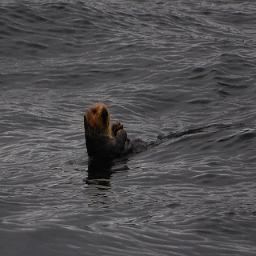} & 
		\includegraphics[width=\imgw]{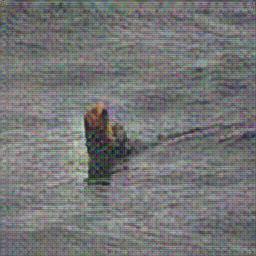} \\
		\includegraphics[width=\imgw]{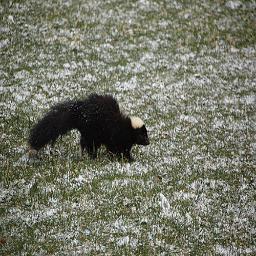} & 
		\includegraphics[width=\imgw]{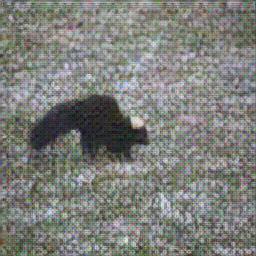} & 
		\includegraphics[width=\imgw]{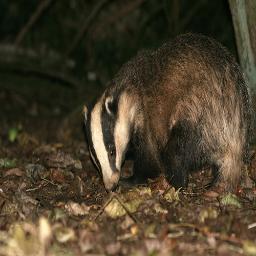} & 
		\includegraphics[width=\imgw]{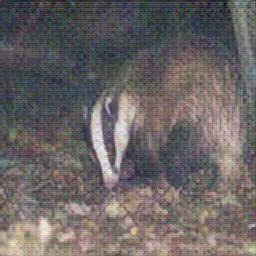} & 
		\includegraphics[width=\imgw]{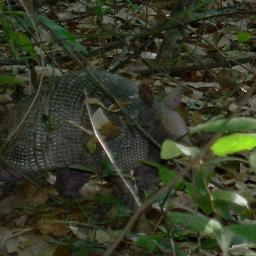} & 
		\includegraphics[width=\imgw]{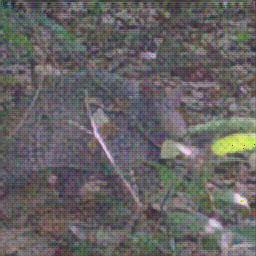} \\
		\includegraphics[width=\imgw]{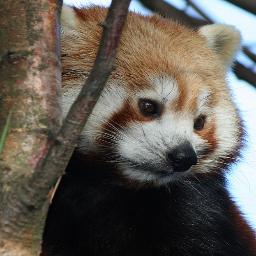} & 
		\includegraphics[width=\imgw]{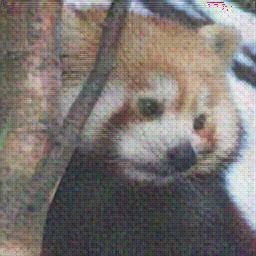} & 
		\includegraphics[width=\imgw]{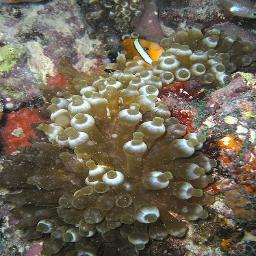} & 
		\includegraphics[width=\imgw]{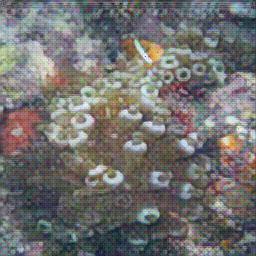} & 
		\includegraphics[width=\imgw]{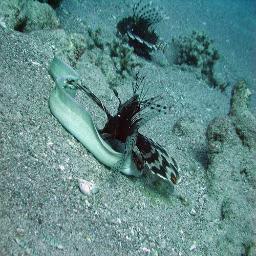} & 
		\includegraphics[width=\imgw]{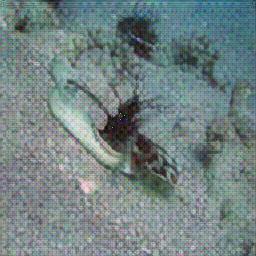} \\
		\includegraphics[width=\imgw]{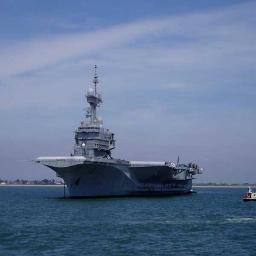} & 
		\includegraphics[width=\imgw]{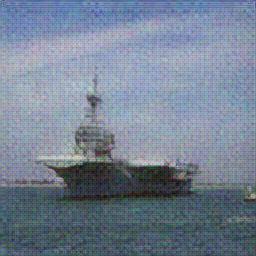} & 
		\includegraphics[width=\imgw]{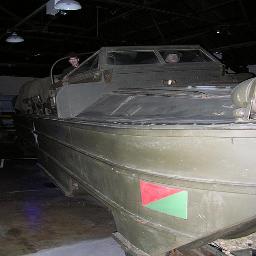} & 
		\includegraphics[width=\imgw]{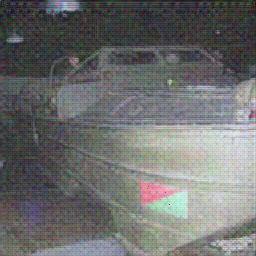} & 
		\includegraphics[width=\imgw]{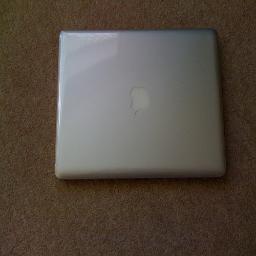} & 
		\includegraphics[width=\imgw]{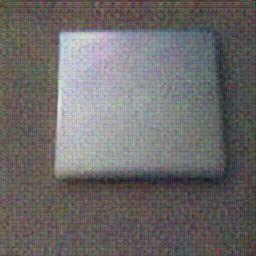} \\
		\includegraphics[width=\imgw]{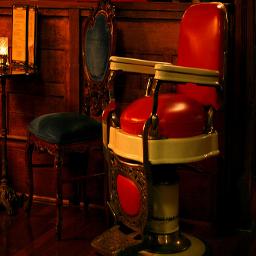} & 
		\includegraphics[width=\imgw]{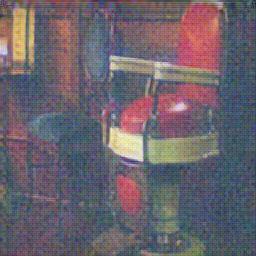} & 
		\includegraphics[width=\imgw]{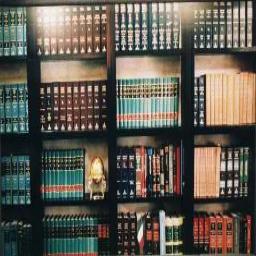} & 
		\includegraphics[width=\imgw]{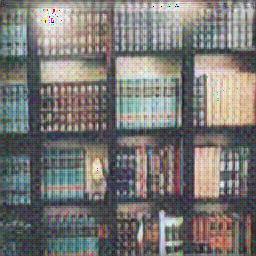} & 
		\includegraphics[width=\imgw]{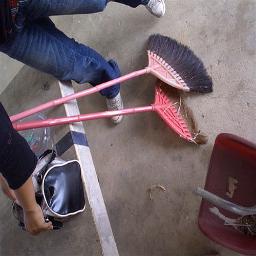} & 
		\includegraphics[width=\imgw]{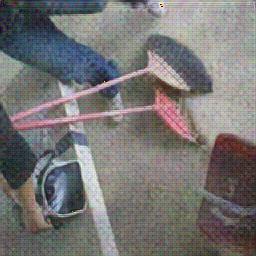} \\
		\includegraphics[width=\imgw]{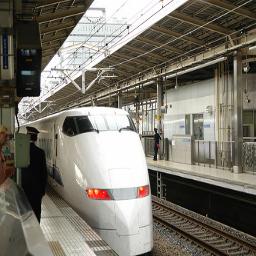} & 
		\includegraphics[width=\imgw]{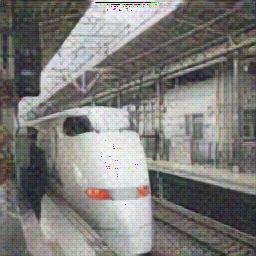} & 
		\includegraphics[width=\imgw]{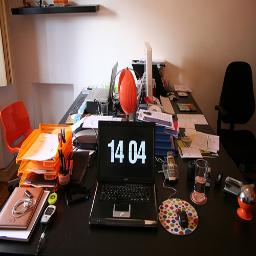} & 
		\includegraphics[width=\imgw]{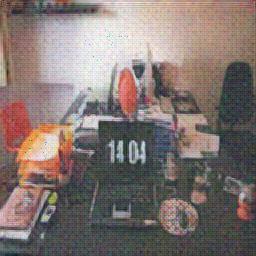} & 
		\includegraphics[width=\imgw]{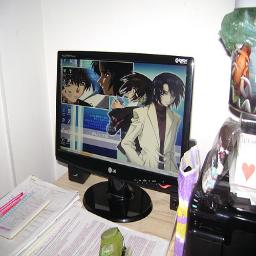} & 
		\includegraphics[width=\imgw]{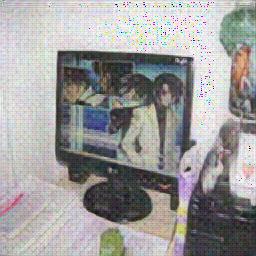} \\
\end{tabular}
\caption{Autoencoder on the ImageNet dataset \cite{ILSVRC15}. Original sample on the left and reconstruction on the right. The images were resized to 256x256 to fit the model. This use the amplitude encoder as circuit are too long to simulate}
\label{fig:imagnet}
\end{figure}

Visually, the reconstructions demonstrate that the amplitude encoder preserves substantial image detail across datasets, as shown for MNIST and CIFAR-10 in Figure~\ref{fig:VAECIFARMNIST} and for ImageNet in Figure~\ref{fig:imagnet}. The autoencoder can encode even the CIFAR-10 dataset into small quantum states and recover the information, suggesting that high-dimensional classical datasets can be encoded into compact quantum representations suitable for small-scale quantum models. The metrics in Table~\ref{tab:reconstruction_metrics} indicate a noisy reconstruction, consistent with a model that prioritizes data compression over pixel-perfect reconstruction.

\begin{figure}[p]
\centering       
\begin{tabular}{ccc}
\includegraphics[scale=0.45]{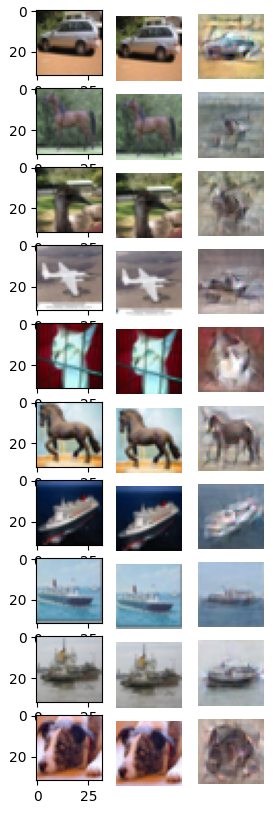} & 
\includegraphics[scale=0.45]{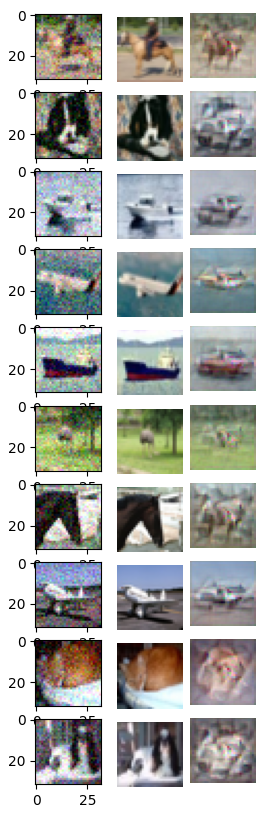}
\includegraphics[scale=0.45]{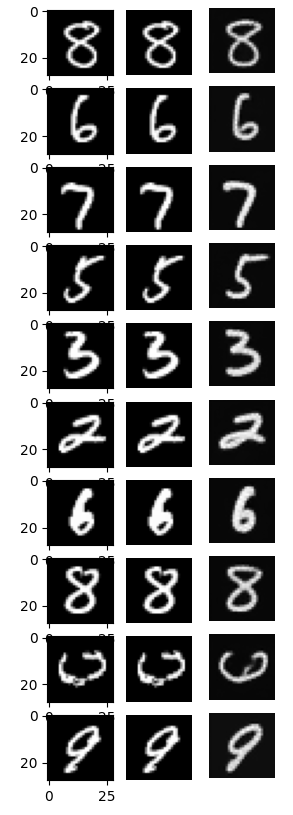}       
\end{tabular}
\caption{
Autoencoder results for the data re-uploading model on the CIFAR dataset. Validation set reconstructions are shown on the left and test set reconstructions on the right. In each case, the first image is the noisy input. The reconstructed CIFAR images are noticeably blurry, suggesting that the quantum circuits may not have sufficient expressive capacity to fully capture the underlying image features.
} 

\label{fig:VAEPRETRAINCIFARMNIST}
\end{figure}

The ansatz encoder shows a different pattern: as shown in Figure~\ref{fig:VAEPRETRAINCIFARMNIST}, its CIFAR-10 reconstructions are noticeably blurrier than those of the amplitude encoder, while its MNIST reconstructions remain sharp. We discuss possible reasons for this gap in Section~\ref{subsec:disscaling}.


For higher dimensional 256x256 resolution images, some blurriness is observed in the reconstruction, as seen in the ImageNet result in Figure \ref{fig:imagnet}. This suggests a lack of capacity to handle all the finer details and eliminate the blurriness. 
For simulated and small circuits mode, it is currently not feasible to embed large image data using the current embedding technique due to the size of the data image. Amplitude embedding requires an input that is too large, even with exponential dimension reduction. Angle embedding does not have exponential dimension reduction, making it even more unfeasible. Data reuploading \cite{P_rez_Salinas_2020} is a more promising technique as it does not require extra circuit width; however, it is limited by our current circuit depth. Current circuits are shallow due to noise and coherence time.

\section{Benchmarking Results}
\label{sec:bench_results}

For the circuit-centric classifier, adding our learned autoencoder 
embedding improves accuracy from 54.0\% with naive amplitude embedding to 
85.0\%. This remains below the 98.5\% achieved by a 2-layer MLP trained on 
the same encoded representations. This gap is consistent with 
\citeauthor{bowles2024betterclassicalsubtleart}~\cite{bowles2024betterclassicalsubtleart}, 
who find the circuit-centric classifier to be the weakest architecture 
among those they survey and attribute part of this weakness to the 
amplitude embedding used to load classical data into the circuit. While 
our learned embedding substantially improves on the naive amplitude 
baseline, the remaining gap to the classical MLP suggests that a 10-qubit 
circuit-centric ansatz may also lack the representational capacity 
required for this task.

To close this gap, we follow the dressed quantum circuit approach 
of~\citeauthor{Mari_2020}~\cite{Mari_2020}, in which a small classical 
network is placed after the quantum measurement outputs. This is a 
concrete instance of the offloading strategy outlined in 
Section~\ref{subsec:Motivation}: rather than requiring the quantum circuit 
alone to perform the full classification task, we let it produce a 
representation that a minimal classical component can then exploit. 
Adding a post-measurement softmax layer, consisting of just two neurons 
and receiving the expectation value of each qubit, raises accuracy from 
85.0\% to over 98.5\%, closing nearly all of the remaining gap to the 
classical MLP. This layer improves optimization by converting the 
measurements into normalized class probabilities compatible with 
cross-entropy loss, yielding stronger and more stable gradients during 
training~\cite{Goodfellow-et-al-2016}.

This large improvement raises the question of how much of the final 
98.5\% accuracy is actually attributable to the quantum circuit, as 
opposed to the two-neuron classical post-processing network. To isolate 
this contribution, we evaluate a baseline in which the quantum circuit is 
removed entirely: the encoder's expectation values are fed directly into 
the same two-neuron classifier, with the autoencoder weights held fixed. 
This baseline reaches only 66.5\%, more than 30 percentage points below 
the full pipeline. This confirms that the quantum circuit performs a 
nontrivial transformation of the encoded features, producing a 
representation that the classical post-processing network can exploit far 
more effectively than the raw encoder output.

\begin{figure}[t]
\centering
V\includegraphics[width=0.75\linewidth]{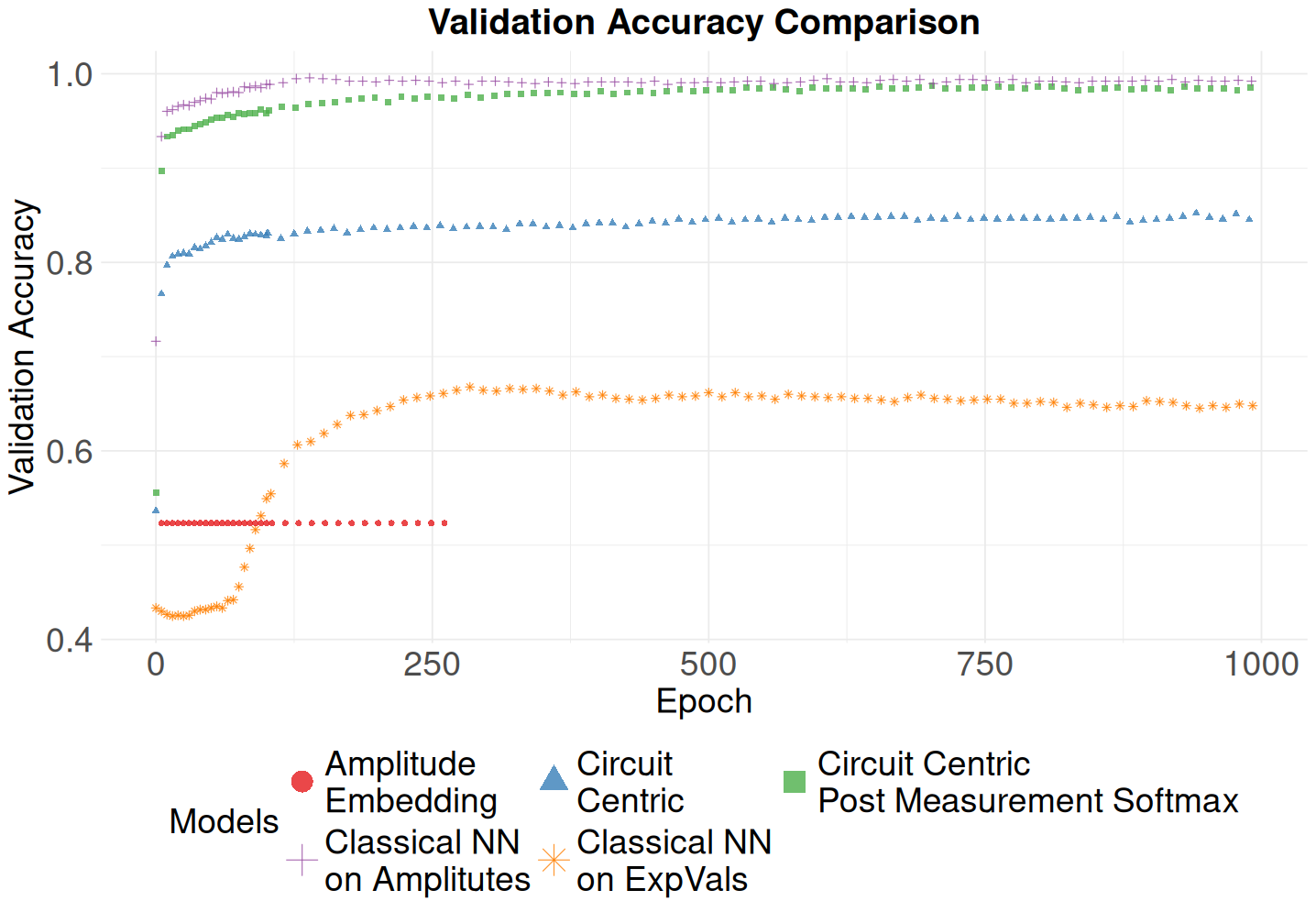}
\caption{Accuracy on the validation dataset of 4 different models on 3v5 MNIST data. The circuit centrist model with amplitude embeddings (red circle) does not converge and so was stopped at 255 epochs. The autoencoder allows the circuit-centric model to reach $0.85$ accuracy  (blue triangle). Adding post-measurement softmax lets the circuit come under $0.05$ accuracy (green square)of a classical neural network performance on the amplitude (purple cross). To test that the quantum circuit is used, we trained a neural network on the expectation value of the encoded states (orange stars), it underperformed by around $30\%$.}

\label{fig:valloss10qb}
\end{figure}

Figure~\ref{fig:valloss10qb} summarizes these results on the 3-vs-5 MNIST 
validation set. The circuit-centric model with naive amplitude embeddings 
(red circle) fails to converge and is stopped at 255 epochs. Introducing 
our learned autoencoder embedding allows the same architecture to reach 
$0.85$ accuracy (blue triangle). Adding the post-measurement softmax layer 
(green square) brings performance to within $0.05$ of the classical MLP 
trained on the same embedding (purple cross). The orange stars show the 
no-quantum-circuit ablation, which underperforms the full pipeline by 
approximately 30 percentage points, confirming that the quantum circuit is 
responsible for this gain.

\begin{figure}[t]
\centering
\includegraphics[width=0.75\linewidth]{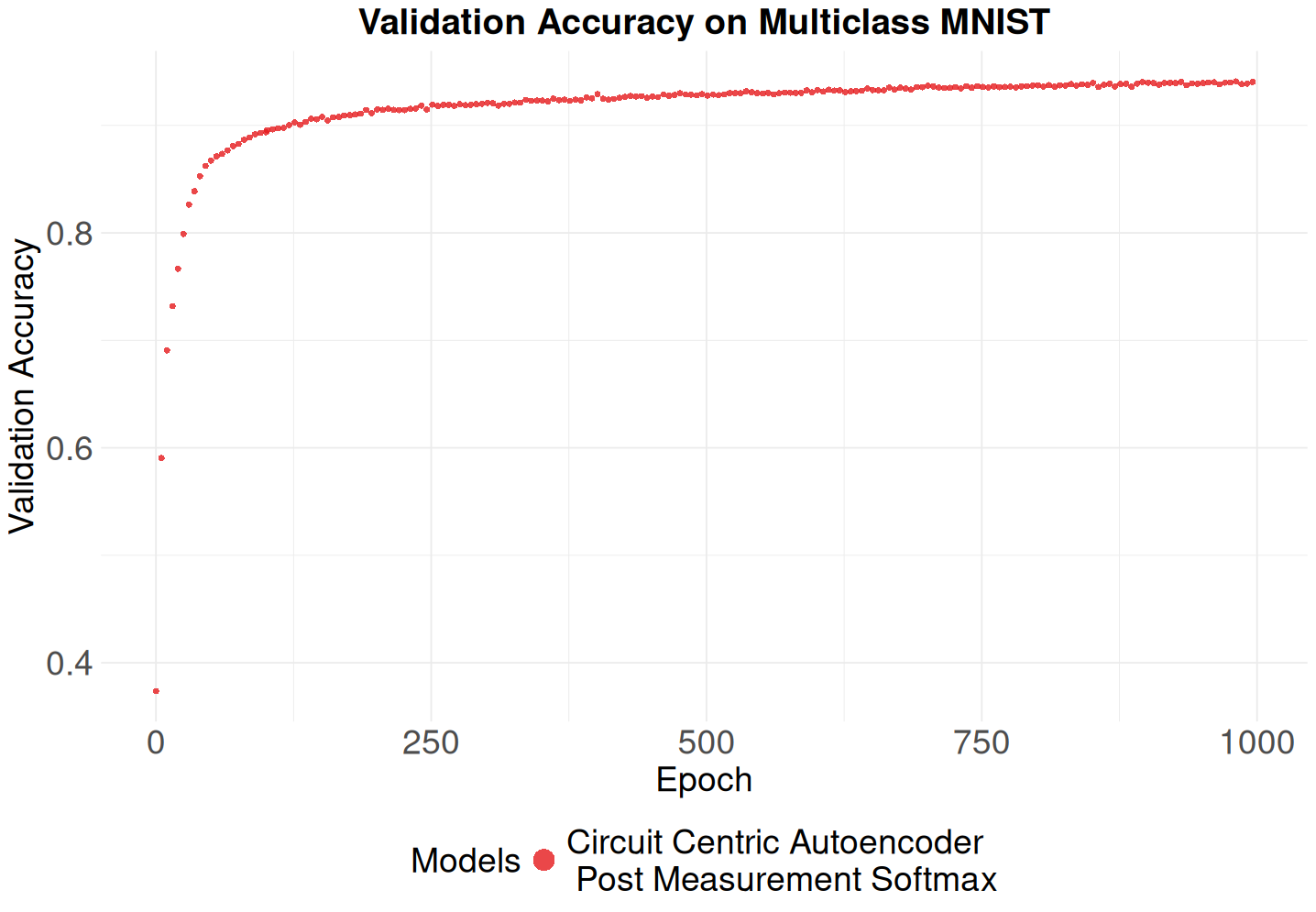}
\caption{Accuracy on the validation dataset of the 10 MNIST classes. This is the performance of the green square model from Figure~\ref{fig:valloss10qb} in multiclass. At 94\% validation, it is 5\% under the classical state of the art \cite{deng2012mnist}.}

\label{fig:valloss10qbMULTI}
\end{figure}

\begin{figure}[t]
\centering
\includegraphics[width=0.75\linewidth]{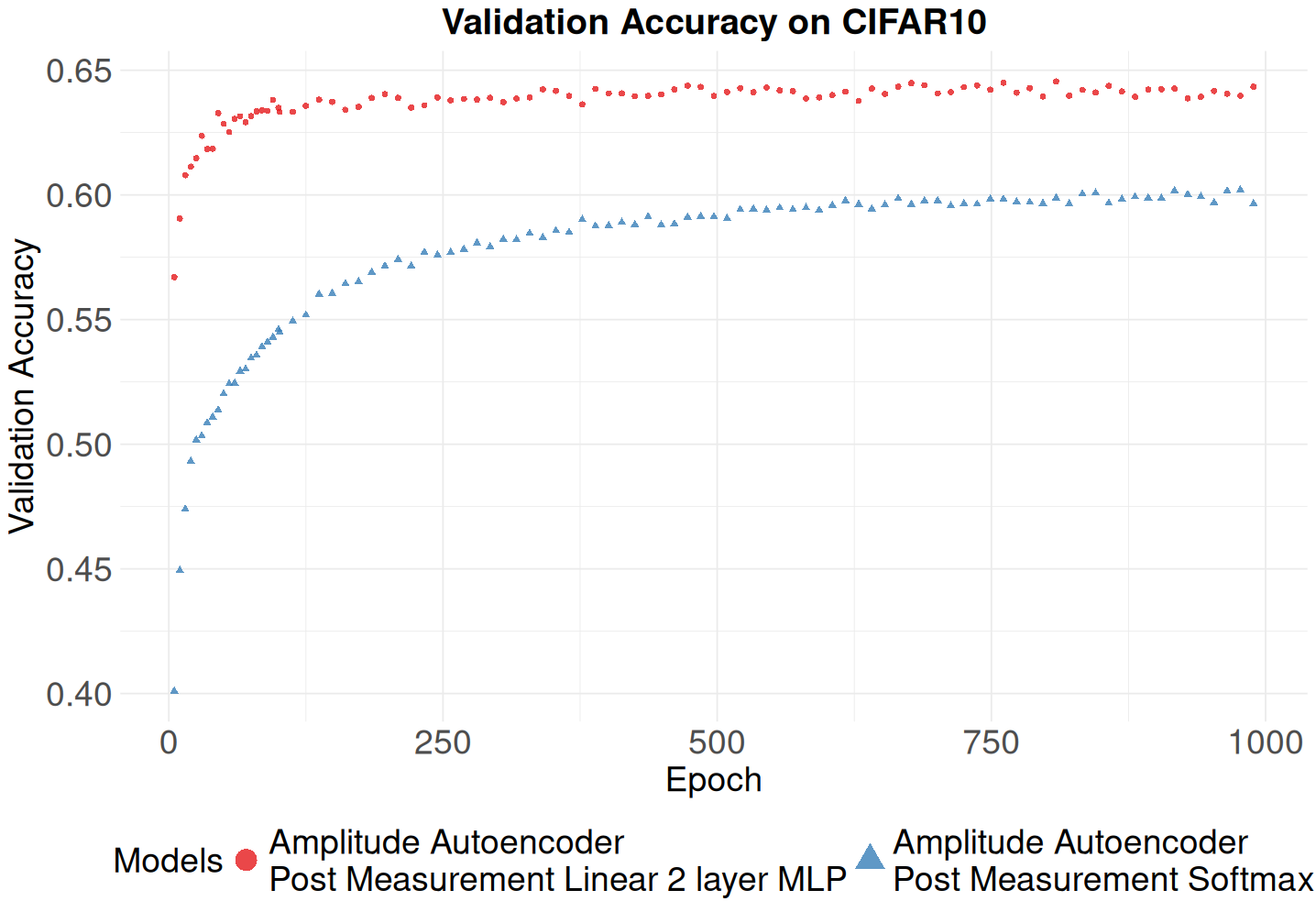}
\caption{Two quantum classifiers. In red, the post-selection softmax was enhanced by two extra linear layers. Our limited simulation speeds and memory usage prevent us, at the moment, from simulating higher capacity circuits. Which we believe would perform much better. }

\label{fig:vallosscifar}
\end{figure}

On CIFAR-10, the circuit-centric classifier with the learned amplitude embedding reaches approximately 60\% validation accuracy, shown in Figure~\ref{fig:vallosscifar}. Adding a single extra linear layer after the post-measurement softmax raises this to approximately 65\%. As with the MNIST results, this indicates that the bottleneck is not primarily the embedding itself but the representational capacity of the circuit that can be simulated at this scale: a small increase in classical post-processing capacity yields a disproportionately large gain in accuracy. This is consistent with the discussion in Section~\ref{subsec:disscaling}, where simulation cost restricts us to circuits too shallow to fully exploit the information present in the learned embedding. Having shown the binary results we exted the to multiclass case in Figure \ref{fig:valloss10qbMULTI} achieving $94\%$ which is ~5\% under the classical state of the art. 

\begin{figure}[t]
\centering
\includegraphics[width=0.75\linewidth]{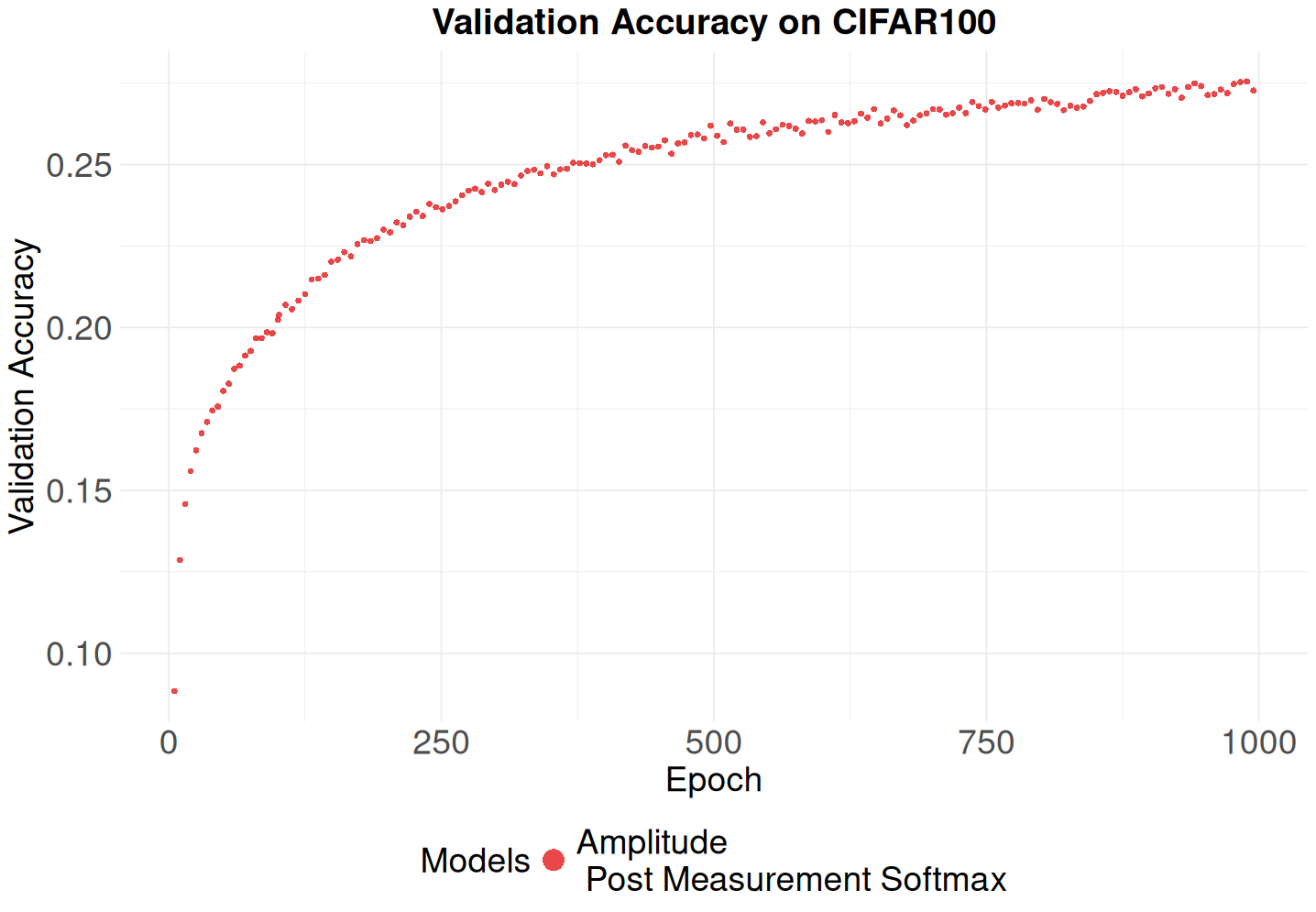}
\caption{Classification on CIFAR-100. The autoencoder used was trained on CIFAR-10. We at $27.5\%$ validation accuracy we are $7.5\%$ away from the Classical AlexNet baseline \cite{Cifar100ClasssicalComp} }

\label{fig:vallosscifar100}
\end{figure}

To test whether the learned embedding generalizes beyond its training distribution, we evaluate the same CIFAR-10-trained autoencoder on CIFAR-100, without retraining the encoder. The resulting classifier reaches approximately 27.5\% validation accuracy (Figure~\ref{fig:vallosscifar100}), within 7.5\% of the classical AlexNet baseline~\cite{Cifar100ClasssicalComp}. Since the autoencoder was never exposed to CIFAR-100 images during training, this result indicates that the embedding captures features general enough to support classification on a related but unseen 100-class distribution, rather than overfitting to the specific classes seen during autoencoder training.

Taken together, these results show that a learned, recoverable embedding 
substantially narrows the gap between quantum and classical models on this 
task, and that offloading the final classification step to a minimal 
classical post-processing network is sufficient to realize most of the 
remaining gain, provided the quantum circuit itself contributes a 
non-trivial transformation. This raises a natural question: why does the 
circuit-centric architecture require this additional offloading step to 
approach classical performance at all, and what does this imply about the 
division of labor between embedding, circuit, and classical post-processing 
in determining quantum model performance? We return to this question in 
Section~\ref{subsec:outlook}.

\section{Discussion}
 \label{sec:discussion}


The results presented above show that a learned quantum embedding substantially narrows the gap between quantum and classical models. However, interpreting these gains requires care: quantum machine learning models are known to underperform classical baselines~\cite{bowles2024betterclassicalsubtleart}, and reported advantages can be sensitive to evaluation choices. We therefore discuss the robustness of our comparisons, the scalability of the two encoders, and the broader engineering challenges that remain.

\subsection{Robustness of Reported Results}

Interpreting the performance of quantum machine learning models requires careful consideration of the comparison protocol used. Reported quantum advantages can sometimes arise from differences in model tuning or evaluation procedures rather than from the quantum model itself.
In this context of 
\citeauthor{bowles2024betterclassicalsubtleart}~\cite{bowles2024betterclassicalsubtleart}, 
showed that quantum classifiers are consistently outperformed by 
classical baselines across 160 datasets. They further caution that 
heavily tuned quantum models compared against untuned classical baselines 
can produce misleading performance advantages. To avoid this, we 
perform only minimal tuning by selecting the learning rate and 
retaining the best checkpoint by validation loss, and compare 
against a classical MLP trained on the same encoded representations 
under identical conditions. The 98.5\% we report is therefore a 
conservative estimate of what our embedding can achieve, and the 
remaining 1.2\% gap to the classical MLP (99.7\%) reflects circuit 
capacity constraints rather than embedding quality.

The amplitude embedding baseline of
\citeauthor{bowles2024betterclassicalsubtleart}~\cite{bowles2024betterclassicalsubtleart} 
relies on PCA for dimensionality reduction prior to embedding. In 
contrast, our raw amplitude baseline uses no such preprocessing and 
fails to converge entirely, confirming that PCA is doing non-trivial 
work in their pipeline. Our autoencoder replaces PCA with a learned 
embedding jointly optimized for quantum compatibility, achieving 98.5\%. Having established that the reported gains reflect the embedding itself rather than artifacts of tuning or preprocessing, we now discuss what these results contribute relative to prior work. 

\subsection{Novelty}

Within current simulation budgets, the amplitude encoder's lower simulation cost made it the practical choice for the largest dataset considered in this work; training the ansatz encoder at ImageNet scale remains an open challenge, though we remain hopeful that improvements in circuit depth and simulation efficiency will make it feasible in the future.
To the best of our knowledge, this is the first demonstration of 
quantum embedding and reconstruction on ImageNet (Figure~\ref{fig:imagnet}), previously 
considered intractable with standard embeddings: amplitude embedding 
requires exponential state preparation cost at that scale, while 
angle embedding requires a number of qubits linear in the number of 
features, making 256$\times$256$\times$3 images entirely out of reach 
for both approaches without dimensionality reduction. Notably, this result was achieved after only 5 epochs of training due to the substantial computational cost, with training taking several weeks to complete, suggesting that reconstruction quality represents a lower bound on what the framework can achieve with further training. Our embedding framework is agnostic to the choice of quantum 
circuit, measurement strategy, and dataset, making it a 
general-purpose tool rather than a task-specific solution. Unlike hybrid architectures that jointly train classical and quantum components~\cite{bowles2024betterclassicalsubtleart,Liu_2025, Mari_2020, }, our approach decouples the autoencoder from the quantum classifier by fixing the encoder weights before classification training, allowing the quantum circuit's contribution to be evaluated independently rather than being absorbed into a jointly trained end-to-end pipeline.

\subsection{Scaling and Hardware Considerations}
\label{subsec:disscaling}

The amplitude encoder, while effective in simulation, requires a 
classical description of the full quantum state, which scales 
exponentially in the number of qubits. It is therefore limited to 
small circuits and cannot be directly executed on quantum hardware 
without an efficient state preparation routine. One promising direction 
to address this is to stack multiple smaller tractable state descriptions 
as separable states, effectively enabling amplitude encoding at larger 
scales without requiring an exponentially large classical description. 
This comes at the cost of long-range entanglement across the full state, 
though it has the advantage of avoiding barren plateaus during embedding 
training. We leave a systematic study of this trade-off as future work.

The ansatz encoder addresses the scaling limitation more directly: by 
mapping the learned latent variables to rotation angles of single-qubit 
gates, the memory and runtime cost scale polynomially with the number 
of qubits, making it compatible with both current NISQ devices and 
future fault-tolerant hardware. This was confirmed by executing the 
ansatz encoder on IBM quantum hardware in inference mode, where the 
learned representations remained stable and reconstructable under real 
device noise (Figure ~\ref{fig:ibm_results}). The trade-off is that quantum simulation of the ansatz 
encoder during training is significantly slower than the amplitude 
encoder, as it requires explicit circuit simulation rather than linear 
algebra. As quantum software and hardware mature, the ansatz encoder 
is the more promising path toward practical large-scale deployment. 

Beyond scalability, the two encoders also differ in reconstruction quality at comparable or larger qubit counts. As shown in Figure~\ref{fig:VAEPRETRAINCIFARMNIST}, the ansatz encoder (10 qubits) reconstructs MNIST sharply but produces noticeably blurry CIFAR-10 reconstructions, despite using more qubits than the amplitude encoder (5 qubits for MNIST, 7 for CIFAR-10; Figure~\ref{fig:VAECIFARMNIST}), which reconstructs CIFAR-10 with much higher fidelity. This suggests that expressivity is limited by circuit depth rather than qubit count. Scaling the ansatz encoder to harder datasets will therefore require deeper circuits, though this brings its own challenges: greater depth increases sensitivity to the chosen circuit topology and raises the risk of barren plateaus~\cite{McClean_2018}. Improving the ansatz encoder's performance on harder datasets will likely require greater circuit depth rather than additional qubits. Although, as discussed above, both remain limited by simulation cost.

\subsection{The Road to Quantum Advantage}
\label{subsec:outlook}
Returning to the question raised in Section~\ref{sec:bench_results}: why a circuit-centric architecture requires classical post-processing to approach classical performance at all, we hypothesize this reflects a broader immaturity in quantum machine learning engineering, rather than a fundamental limitation of the circuits themselves.

Let us imagine a model is given to learn a task where quantum has a currently known possible advantage such as factoring. We will call it MNIST-FACTORING. You are given an image of a large number and must output one or more no trivial factors of the number (one factor is enough). 
Now a quantum model could learn to implement Shor's algorithm to solve this problem and obtain an advantage over classical models. There is no guarantee we can learn this algorithm.  
This is an example of a task where we have known speedup over classical algorithms. In reality, currently, state-of-the-art quantum machine learning models are outperformed by classical deep learning and in some cases linear models \cite{bowles2024betterclassicalsubtleart}. 

We hypothesize that the underperformance is due to  deficiencies in machine learning art. Research on hyperparameter selection for quantum models, such as optimizer selection, initial value and input embeddings choice, 
is years behind classical models. Classical deep learning models used to suffer from similar underperformance \cite{glorot2010understanding,he2015delving,Goodfellow-et-al-2016,lecun1998efficient} before many modern techniques were created. 
We know there exist many classical operations for which there will not be any quantum speedups. Quantum machine learning models are either executed on current quantum hardware or simulated. The size of 10--25 qubits in simulation or 10 to 48+ noisy qubits in 
hardware is insufficient for classical operations such as arithmetic, 
for which no quantum speedup is currently 
known~\cite{wang2024comprehensivestudyquantumarithmetic}.

In those cases where qubits are equivalent to bits current machines with giga to terabytes of memory are exponentially more powerful than 50 bits. To gain benefit from current quantum hardware it logically follows that best practices are to offload classical computation out of the quantum hardware. 
The main difficulty when offloading computing to classical hardware is identifying where and if the quantum part of the algorithm is giving us any advantages. Furthermore, current quantum 
computers often operate below 64 qubits, making even 32-bit float 
arithmetic impossible to perform natively on them. 

Whether image classification will ever yield a quantum advantage remains an open question. What our framework establishes is that a poorly chosen embedding is itself a source of underperformance, independent of that question, and that a principled quantum embedding is a necessary first step regardless of where quantum advantage ultimately lies. These open questions notwithstanding, the present results already mark concrete progress toward closing the gap with classical models, as we summarize below.

\section{Conclusion}

In this work, we presented the embedding of large-scale datasets by successfully encoding ImageNet with satisfactory reconstruction quality on
13-qubit circuits. The proposed method is both scalable and generic, and its performance is expected to improve alongside advances in
computational power and quantum hardware. Moreover, the approach is applicable to both current
quantum hardware and future fault-tolerant quantum hardware.
As mentioned in the Discussion, we benchmarked the empirical performance of our approach on CIFAR-10 and MNIST as representative test cases, recovering much of the performance gap relative to standard neural network classifiers. The achievable performance remains constrained by the size of the quantum circuits that can be simulated and the available GPU resources.

Using our two encoding strategies—the state encoder and the circuit encoder—we developed two complementary approaches. The state encoder enables substantially faster simulations but is limited in terms of scalability and physical realism. In contrast, the circuit encoder scales more naturally and is directly applicable to both NISQ and fault-tolerant quantum hardware, although it is considerably more computationally demanding to simulate and train. Scaling the ansatz encoder 
to ImageNet-scale data will likely require progressive circuit depth 
training~\cite{skolik2021layerwise} as a mitigation strategy for 
barren plateaus~\cite{McClean_2018}, though a systematic study remains 
computationally prohibitive with current simulation tools.


The proposed encoder-decoder framework is sufficiently general to accommodate a wide variety of encoding and decoding schemes, provided that convergence can be achieved. Future research will explore alternative encoder and decoder architectures as well as extensions beyond image classification, including tabular and other non-image data modalities. On the architecture side, incorporating full adversarial 
training following~\cite{Rombach_2022_CVPR} is a natural next step to 
improve reconstruction quality. Finally, a quantum 
Vector-Quantized VAE (VQ-VAE)~\cite{oord2017vqvae} represents an 
interesting direction, as the discrete codebook structure maps naturally 
to basis states in quantum computing, potentially enabling a fully 
discrete quantum latent space.


\ack{AL would like to express his sincere gratitude to Professor Kim Tae-Kyun for providing the opportunity to carry out the preliminary work on autoencoders in his class and to Renaud B\'echade and Azaetek inc for providing cloud GPU and QPU. }

\funding{AL and D\v{S} acknowledge the support from the Institute for Basic Science in Korea (IBS-R024-D1). D\v{S} further acknowledges funding from the Czech Science Foundation through the Junior Star grant 25-17250M and from Charles University in Prague through PRIMUS/25/SCI/027.}

\newpage
\printbibliography

\end{document}